\documentclass{kapedbk} % Computer Modern font calls
\usepackage{psfig}

\kluwerbib
\def\sun{\hbox{$\odot$}}

\def\lesssim{\mathrel{\hbox{\rlap{\hbox{\lower4pt\hbox{$\sim$}}}\hbox{$<$}}}}
\def\gtrsim{\mathrel{\hbox{\rlap{\hbox{\lower4pt\hbox{$\sim$}}}\hbox{$>$}}}}

\def\arcmin{\hbox{$^\prime$}}

\let\la=\lesssim                        % For Springer A&A compliance...
\let\ga=\gtrsim                                                                 
\begin{document}
\articletitle{X-Ray Observations of Cluster Mergers}
\vskip-1.5truein
\noindent To appear in {\it Merging Processes in Clusters of Galaxies},
edited by L. Feretti, I. M. Gioia, and G. Giovannini
(Dordrecht: Kluwer), in press (2001)
\vskip0.85truein

\articlesubtitle{Cluster Morphologies and Their Implications}

\author{David A. Buote}
\affil{Department of Physics and Astronomy\\ University of California
at Irvine\\ 4129 Frederick Reines Hall\\ Irvine, CA 92697-4575}
\email{buote@uci.edu}

\begin{abstract}
X-ray observations have played a key role in the study of substructure
and merging in galaxy clusters.  I review the evidence for cluster
substructure and mergers obtained from X-ray observations with
satellites that operated before {\sl Chandra} and {\sl XMM}.
Different techniques to study cluster mergers via X-ray imaging and
spectral data are discussed with an emphasis on the quantitative
analysis of cluster morphologies.  I discuss the implications of
measurements of cluster morphologies for cosmology and the origin of
radio halos.
\end{abstract}


\section{Introduction}

``Substructure'' in a galaxy cluster is defined as multiple peaks in
the cluster surface density on scales larger than the constituent
galaxies; the ``cluster surface density'' refers either to the
galaxies, the X-ray emission from hot gas, or the dark matter.  Today
we take it for granted that many galaxy clusters exhibit substructure
and thus are in early stages of formation. This, of course, was not
always the case. In the 1980s there were several searches for cluster
substructure in the optical, but their results were controversial,
primarily because of the difficulty in assessing the importance of
projection effects and the statistical significance of substructure
(see reviews by West 1990, 1995).

X-ray studies of clusters are less susceptible to contamination from
foreground and background objects than optical studies. The X-ray
luminosity is a strong function of the temperature, or mass, which
means that, e.g., foreground groups contribute proportionally less to
the X-ray emission than they do to the galaxy surface density. X-ray
studies of clusters also have the advantage that the signal is limited
only by the effective area of the detector and exposure time of an
observation whereas optical studies are limited by the finite number
of cluster galaxies.

The reality of substructure in clusters was firmly established with
{\sl ROSAT} observations in the early 1990s. The watershed example is
that of A2256 which had long been thought to be a prototypical relaxed
cluster when examined from the perspective of its galaxy
isopleths. However, in a controversial optical study of A2256,
Fabricant, Kent, and Kurtz (1989) proposed the existence of an
infalling subcluster from analysis of the galaxy velocities even
though they detected no such evidence from the galaxy positions alone.

\begin{figure}[ht]   
\parbox{0.49\textwidth}{
\centerline{\psfig{figure=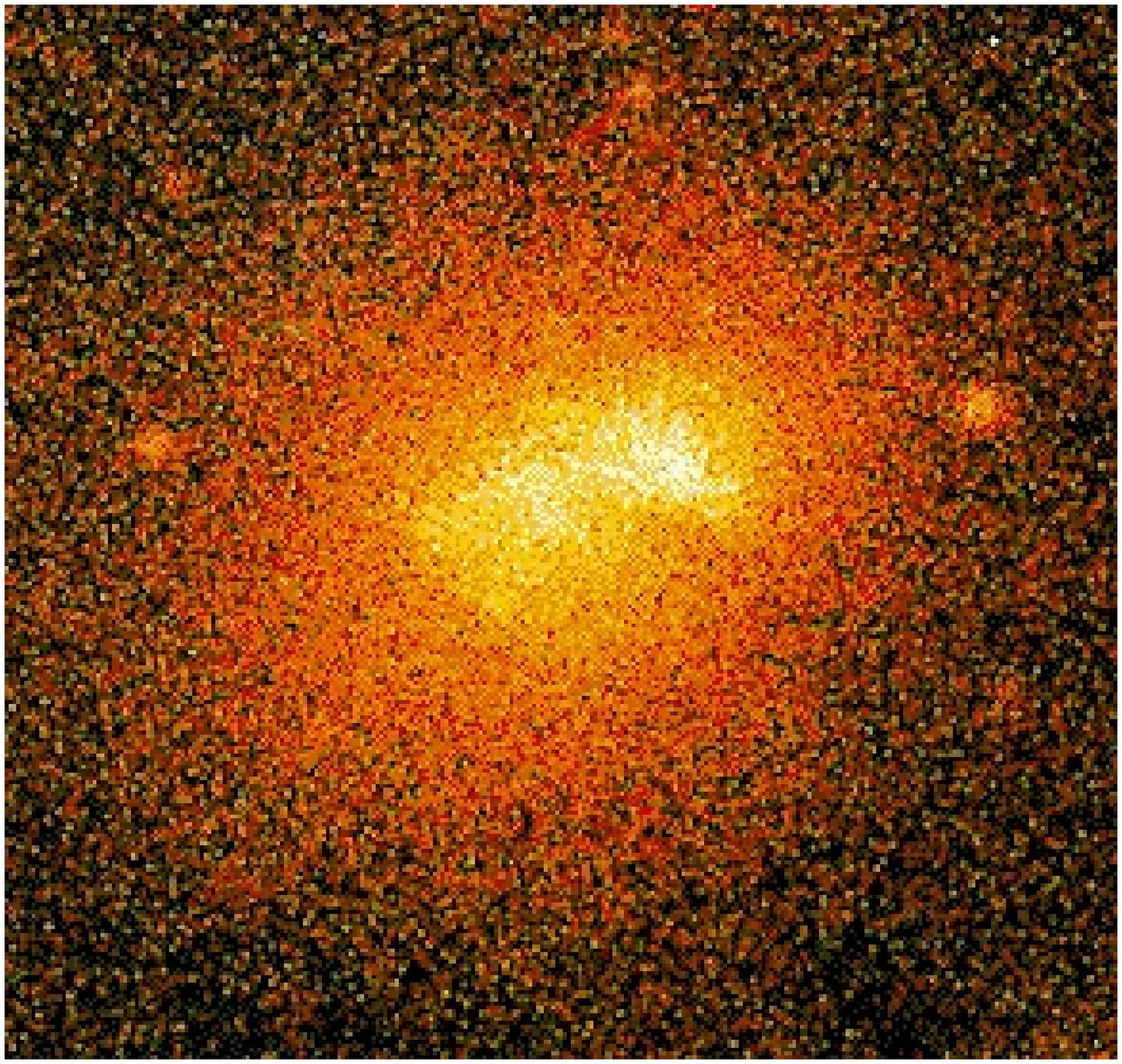,angle=0,height=0.30\textheight}}
}
\parbox{0.49\textwidth}{
\centerline{\psfig{figure=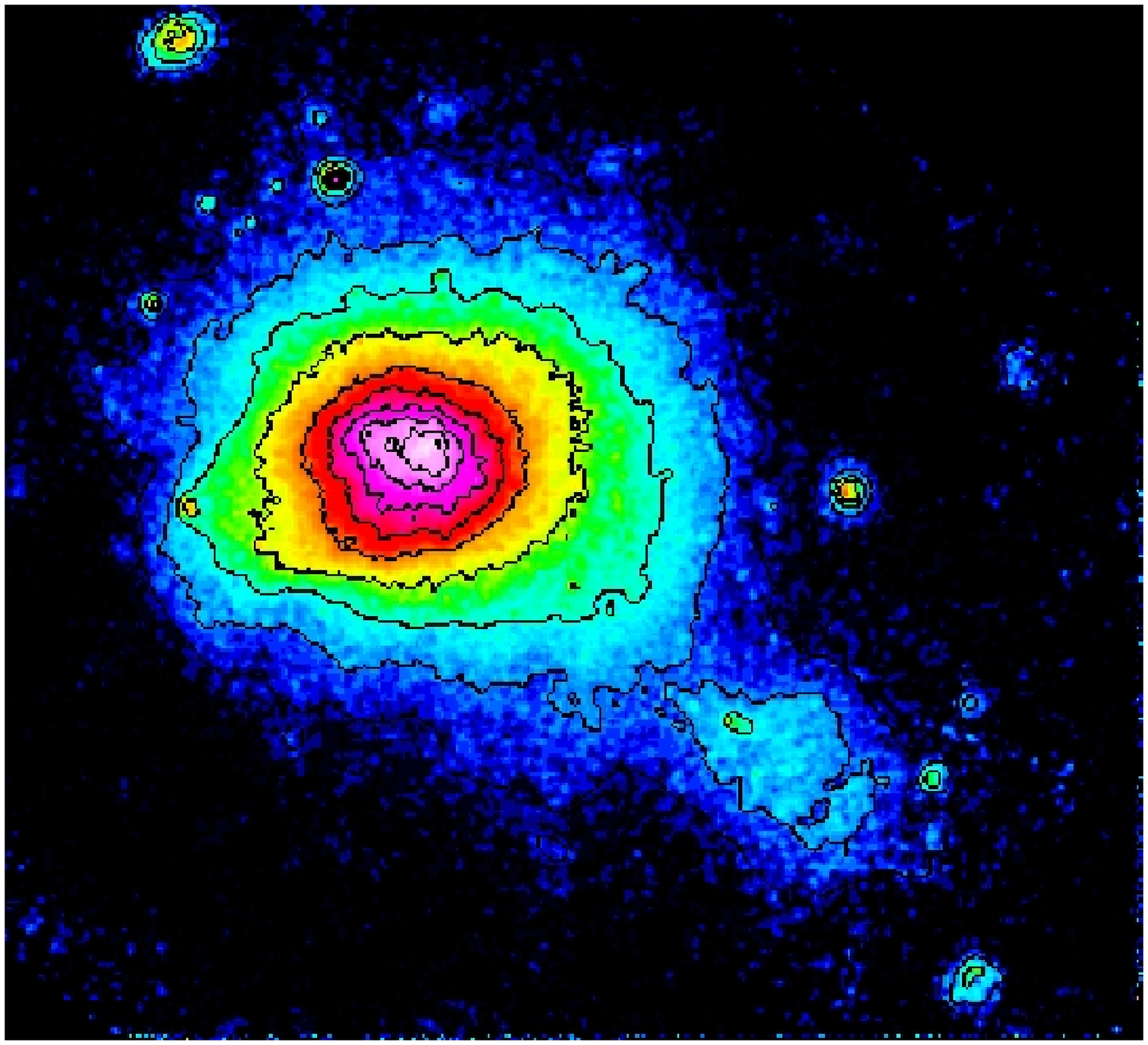,angle=0,height=0.29\textheight}}
}  
\caption{\label{fig.rosat} (Left) {\sl ROSAT} PSPC image of A2256
(Briel et al. 1991). (Right) {\sl ROSAT} PSPC image of Coma (Briel \&
Henry 1997).}
\end{figure}   

The existence of a subcluster in A2256 was confirmed by the stunning
{\sl ROSAT} PSPC image (Briel et al. 1991) that showed a
subcluster\footnote{This X-ray substructure in A2256 could have been
discovered ten years before {\sl ROSAT} since the {\sl Einstein} image
reveals the presence of the subcluster albeit at a lower level of
significance (Buote 1992; Davis \& Mushotzky 1993).}  offset from the
main cluster by a few hundred kpc (Figure
\ref{fig.rosat}, left). {\sl ROSAT} also clearly demonstrated
significant subclustering in the Coma cluster (Briel, Henry, \&
B\"{o}hringer 1992) which had been presumed to be the quintessential
relaxed cluster (Figure \ref{fig.rosat}, right).  Hence, {\sl ROSAT}
images confirmed and clearly established the existence of substructure
in clusters, and thus showed that such clusters are really still
forming.

\begin{figure}[ht]
\parbox{0.49\textwidth}{
\centerline{\psfig{figure=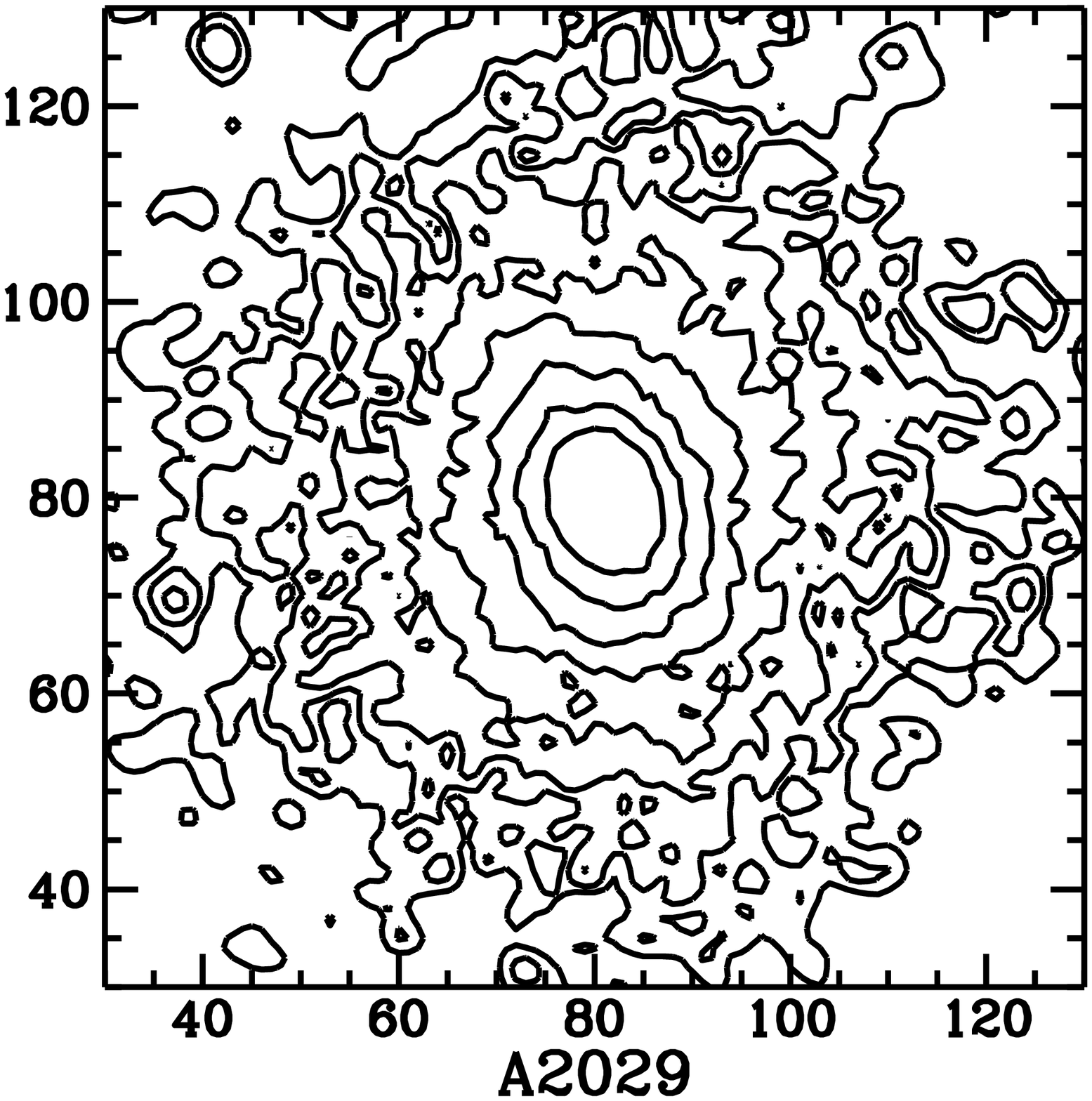,angle=0,height=0.325\textheight}}}
\parbox{0.49\textwidth}{
\centerline{\psfig{figure=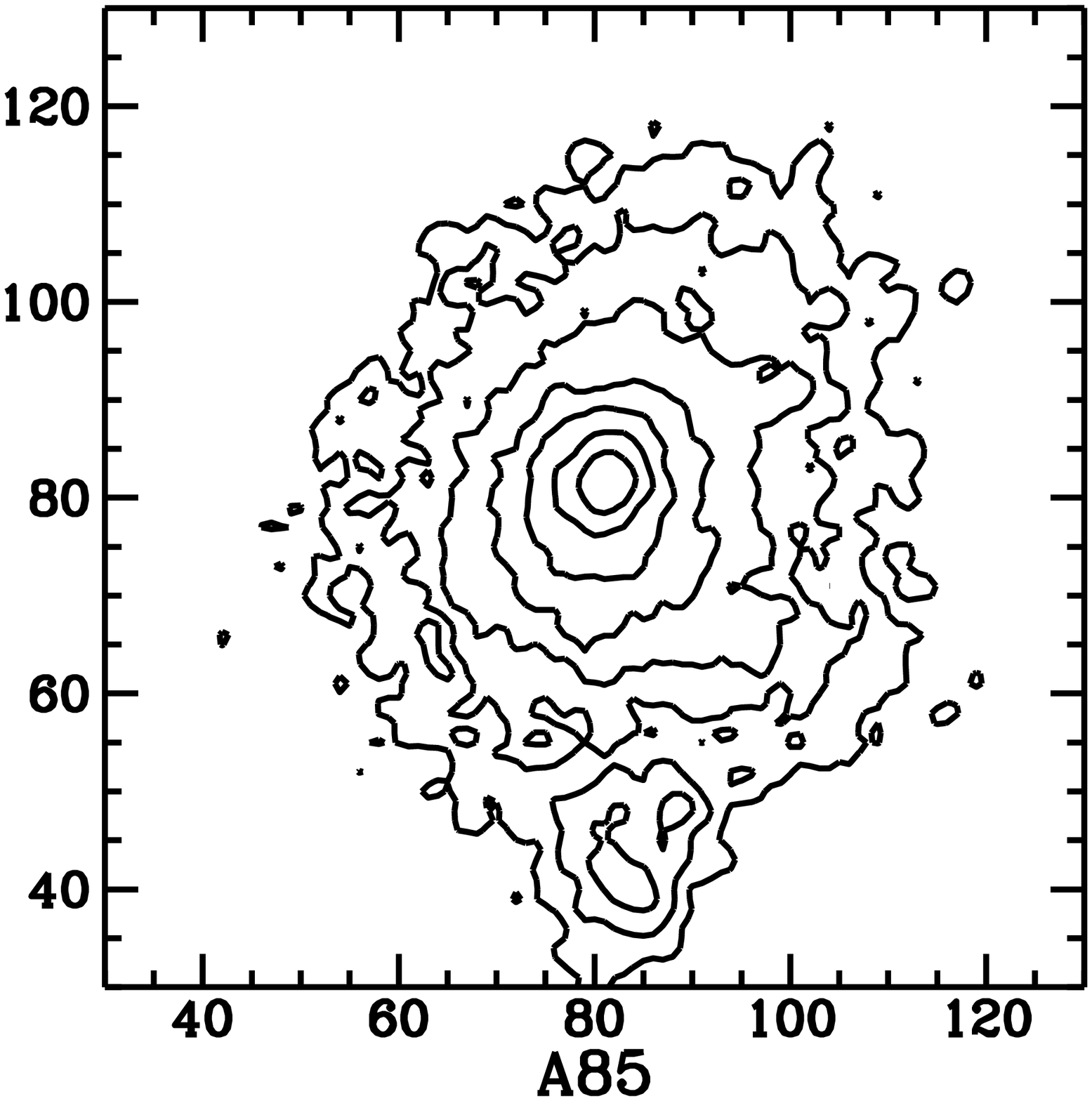,angle=0,height=0.325\textheight}}}
 
\vskip 0.1cm
 
{\large\bf
 
\hskip 2.25cm SINGLE \hskip 3.05cm \large\bf PRIMARY WITH
 
\vskip 0.1cm
 
\hskip 6.85cm SMALL SECONDARY
}
 
\vskip 0.1cm
 
\parbox{0.49\textwidth}{
\centerline{\psfig{figure=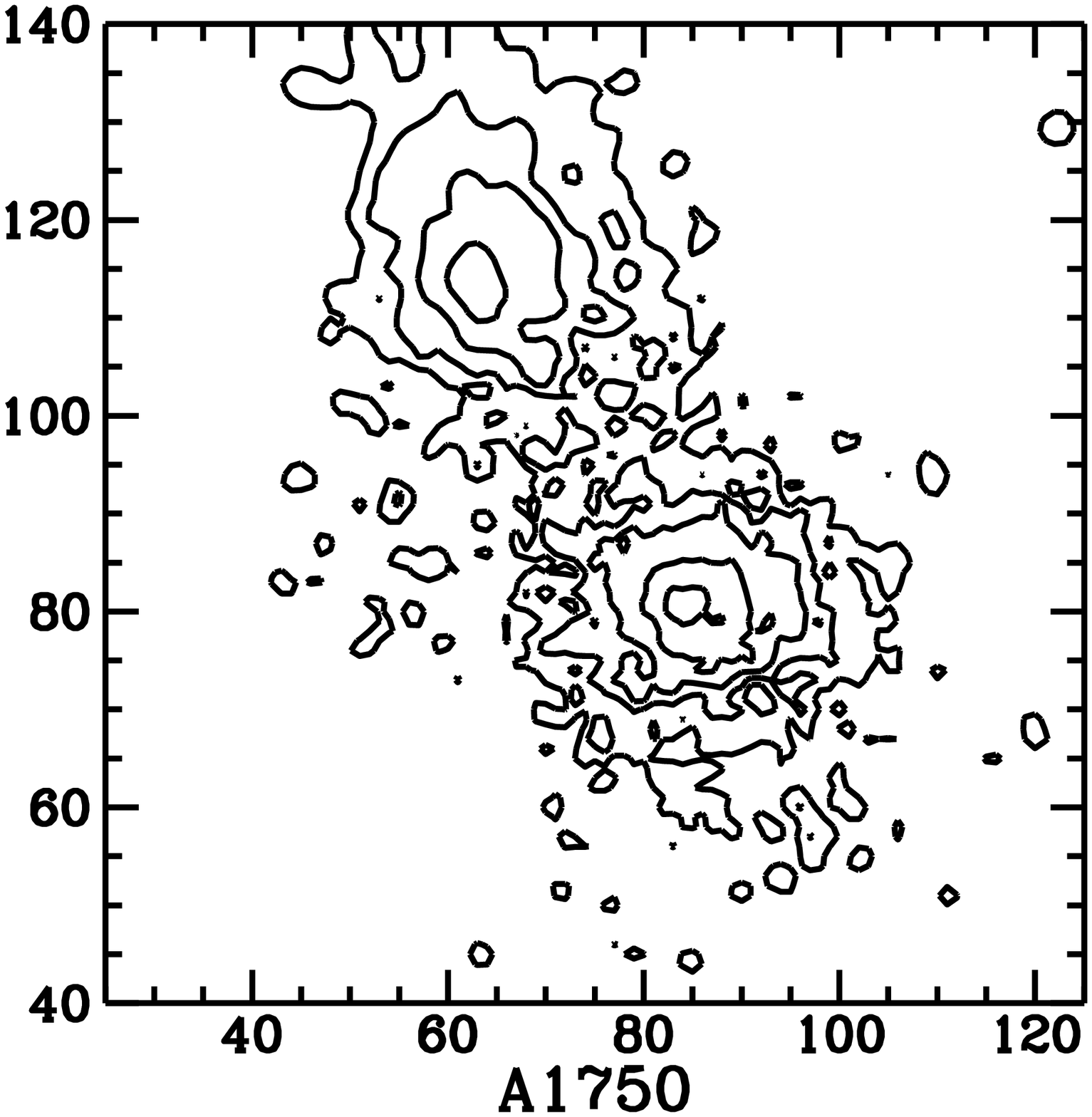,angle=0,height=0.325\textheight}}
}
\parbox{0.49\textwidth}{
\centerline{\psfig{figure=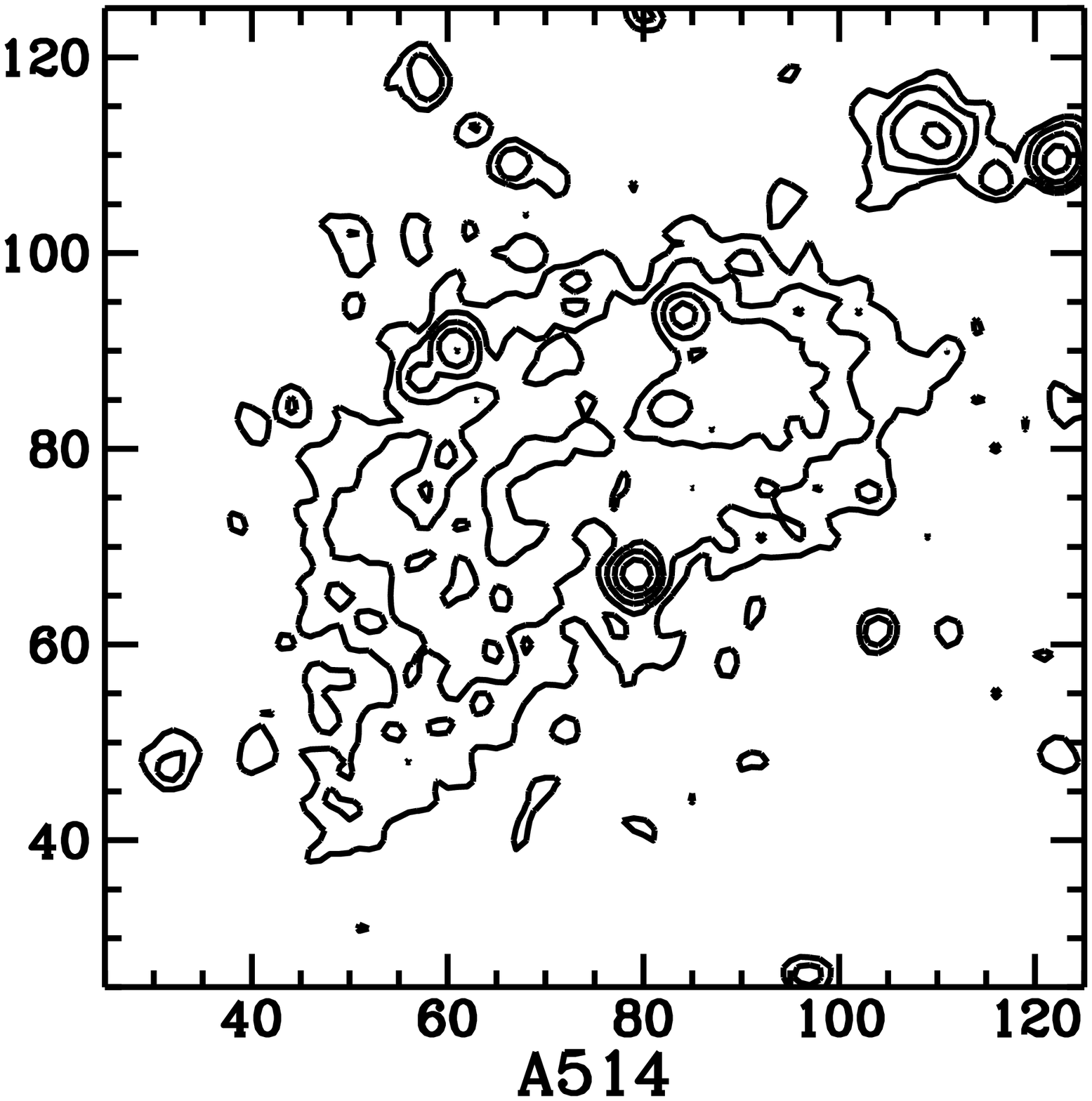,angle=0,height=0.325\textheight}}
}
 
\vskip 0.1cm
 
{\large\bf
 
\hskip 2.1cm DOUBLE \hskip 3.7cm COMPLEX}

\caption{\label{fig.refclusters} Contour plots of {\sl ROSAT} PSPC
images (see Buote \& Tsai 1995) of four Abell clusters labeled
according to the Jones \& Forman (1992) morphological classification
scheme.}

\end{figure}            

The fundamental question raised by these early {\sl ROSAT}
observations is how widespread is merging in clusters? Are clusters
generally young or old? Or is there an equal distribution of cluster
ages in a given cluster sample?  To address this issue one needs to
have measurements of the subclustering properties of a large cluster
sample and, of equal importance, a precise definition of the ``age''
of a cluster. The first systematic X-ray study of cluster merging was
by Jones \& Forman (1992). From visual inspection of $\sim 200$ {\sl
Einstein} cluster images, Jones \& Forman separated the clusters into
6 morphological classes (see Figure
\ref{fig.refclusters}). These classes range from relaxed
single-component systems to systems with a large degree of
substructure. From the relative populations of these classes they
deduced that $\sim 30\%$ of clusters have substructure, which is
actually a lower limit because of the limited resolution of the {\sl
Einstein} IPC. This study established that merging and substructure
are very common in clusters. Consequently, the need arose for a more
precise assignment of the age of a cluster; e.g., how much older or
younger are clusters in the Jones \& Forman classes? Hence, Jones
\& Forman (1992) ushered in the era of quantitative X-ray cluster
morphology.

\section{Quantitative Analysis of Individual Substructures}
\label{detailed}

Quantitative studies of cluster X-ray morphologies have traveled down
two distinctly different paths. The first path is that of the detailed
structural analysis of clusters to determine the number of
substructures, their fluxes, spatial properties, etc.. A popular
approach is to examine the residuals obtained from subtracting a
smooth model representing a relaxed cluster from the X-ray cluster
image (e.g., Davis \& Mushotzky 1993; White et al. 1994; Davis 1994;
Prestwich et al. 1995; Neumann \& B\"{o}hringer 1997, 1999;
B\"{o}hringer et al. 2000). Usually this smooth model is obtained by
fitting a set of perfect elliptical isophotes or an elliptical $\beta$
model to the cluster surface brightness; i.e., the X-ray emission of a
relaxed cluster is assumed to be elliptical in shape.

In hydrostatic equilibrium the surfaces of constant X-ray emissivity
are identical in shape to the surfaces of constant gravitational
potential regardless of the temperature profile of the gas (Buote \&
Canizares 1994, 1998). And since the isopotential surfaces generated
by an elliptical matter distribution (which is assumed to be the most
general stable, relaxed, non-rotating, self-gravitating configuration)
are not perfect ellipses (e.g., Binney \& Tremaine 1987), neither are
the X-ray isophotes. Consequently, the residuals obtained from
subtracting elliptical models from the X-ray surface brightness of
clusters need to be carefully considered.  This procedure is most
appropriately applied as a simple, approximate indicator of
substructure.
 
A more general and powerful method to identify and quantify
substructures is to perform a wavelet decomposition of the X-ray
image. The wavelet analysis is a powerful multi-scale technique to
detect sources embedded in the bright diffuse background cluster
emission which has been successfully applied to many clusters (e.g.,
Slezak et al. 1994; Vikhlinin, Forman, \& Jones 1994; Grebenev et
al. 1995; Biviano et al. 1996; Pislar et al. 1997; Lima-Neto et
al. 1997; Pierre \& Starck 1998; Lemonon et al. 1997; Dantas et
al. 1997; Vrtlik et al. 1997; Lazzati \& Chincarini 1998; Lazzati et
al. 1998; Arnaud et al. 2000). Wavelet analysis locates substructures
on different scales and allows separate spatial analysis (e.g., flux,
extent etc.) of each detected structure. The statistical significance
of the substructures can be assessed rigorously via Monte Carlo
simulations.

\begin{figure}[ht]   
\parbox{0.49\textwidth}{
\centerline{\psfig{figure=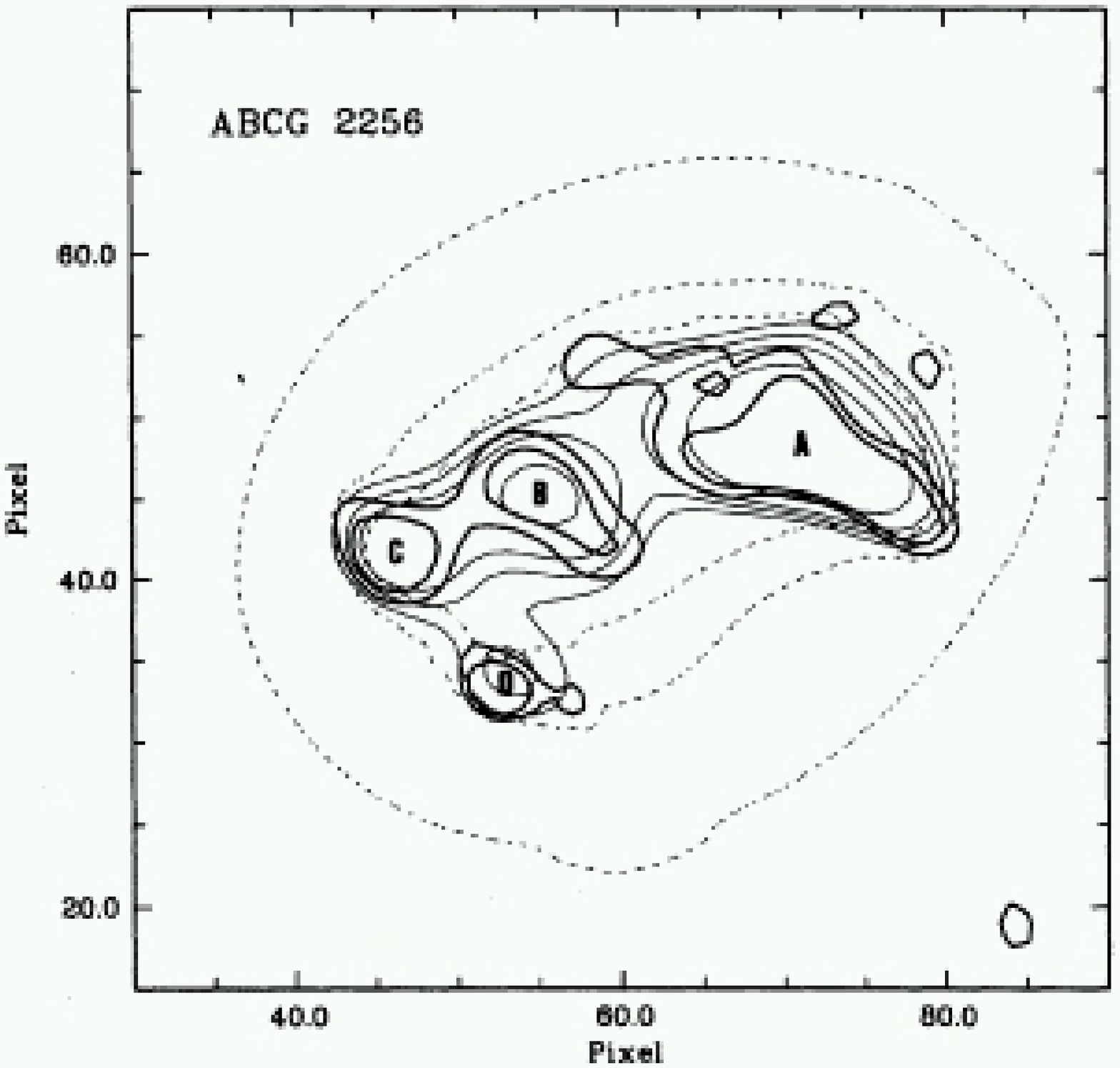,angle=0,height=0.29\textheight}}}
\parbox{0.49\textwidth}{
\centerline{\psfig{figure=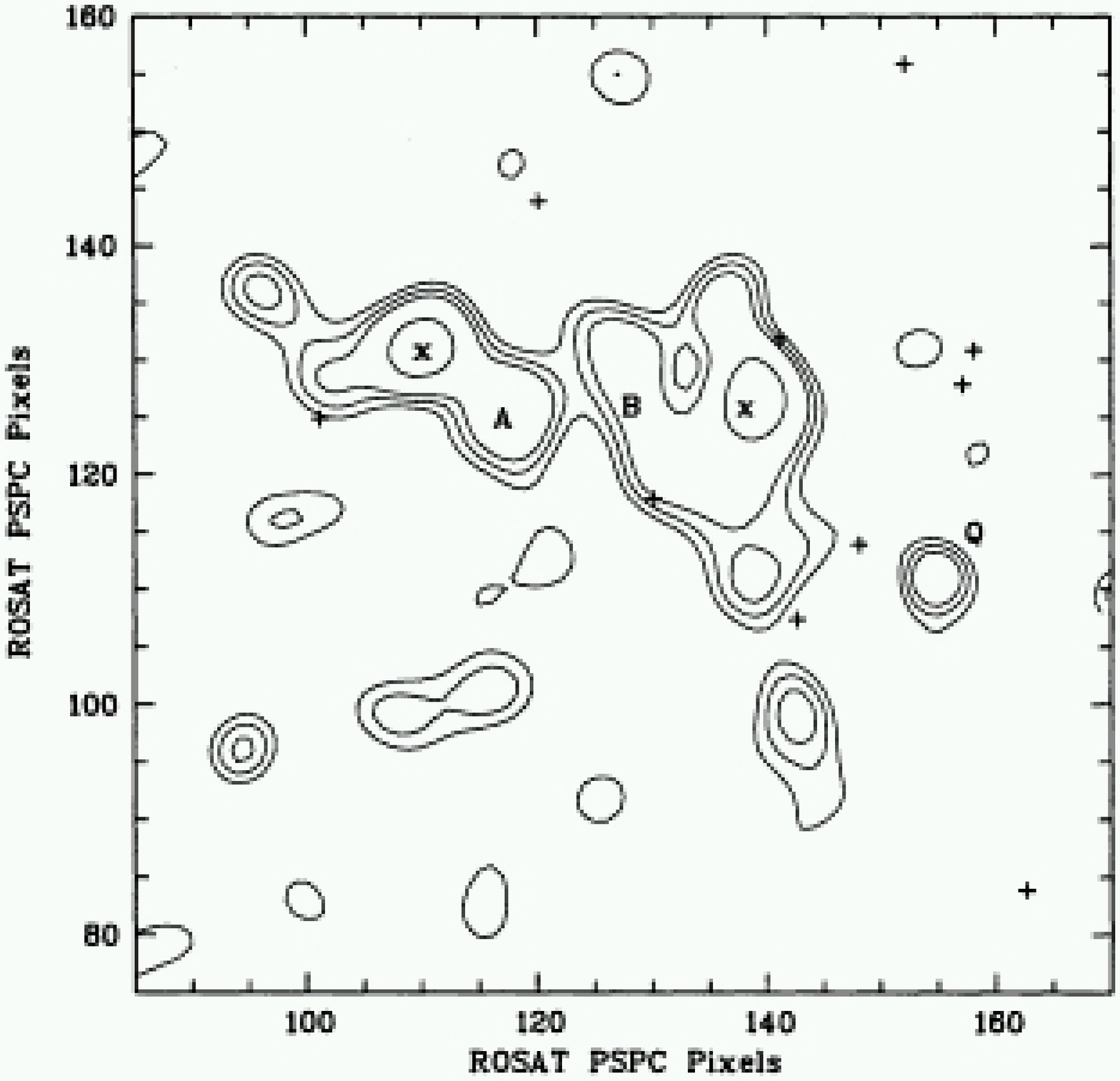,angle=0,height=0.29\textheight}}}   
\caption{\label{fig.wavelet_1} Wavelet decomposition of {\sl ROSAT}
images of (Left) A2256 by Slezak et al. (1994) and (Right) of Coma by
Biviano et al. (1996).}
\end{figure}  

Applications of wavelets to the {\sl ROSAT} images of A2256 and Coma
(Figure \ref{fig.rosat} are shown in Figure \ref{fig.wavelet_1}. In
the case of A2256 Slezak et al. (1994) establish that the core is more
than a simple bimodal system since the bottom-left region consists of
at three subclusters The wavelet analysis of Coma by Biviano et
al. (1996) shows that the core consists of two subclusters surrounding
each of the large galaxies NGC 4874 and NGC 4889. Apparently both Coma
and A2256 are far from relaxed systems.

\begin{figure}[ht]   
\centerline{\psfig{figure=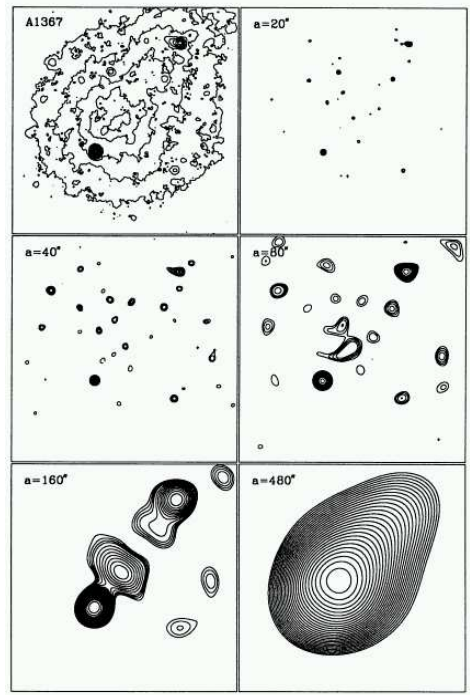,angle=0,height=0.49\textheight}}
\caption{\label{fig.wavelet_2} {\sl ROSAT} image of A1367 and a
wavelet decomposition on different scales by Grebenev et al. (1995). }
\end{figure}  

Wavelets are particularly useful for less-massive systems like A1367
where the emission from several galaxies or groups needs to be
separated from the diffuse cluster background. In Figure
\ref{fig.wavelet_2} is shown the wavelet analysis by Grebenev et al. (1995)
who analyzed both the {\sl ROSAT} PSPC and HRI images and detected 16
extended sources embedded in the diffuse ICM of A1367. Not only does
the wavelet analysis allow the fluxes and extents of each of these
sources to be quantified, but the larger scale wavelets (see Figure
\ref{fig.wavelet_2}) show that the cluster is bimodal with subclusters
centered about what are likely to be galaxy groups.

\begin{figure}[ht]   
\parbox{0.49\textwidth}{
\centerline{\psfig{figure=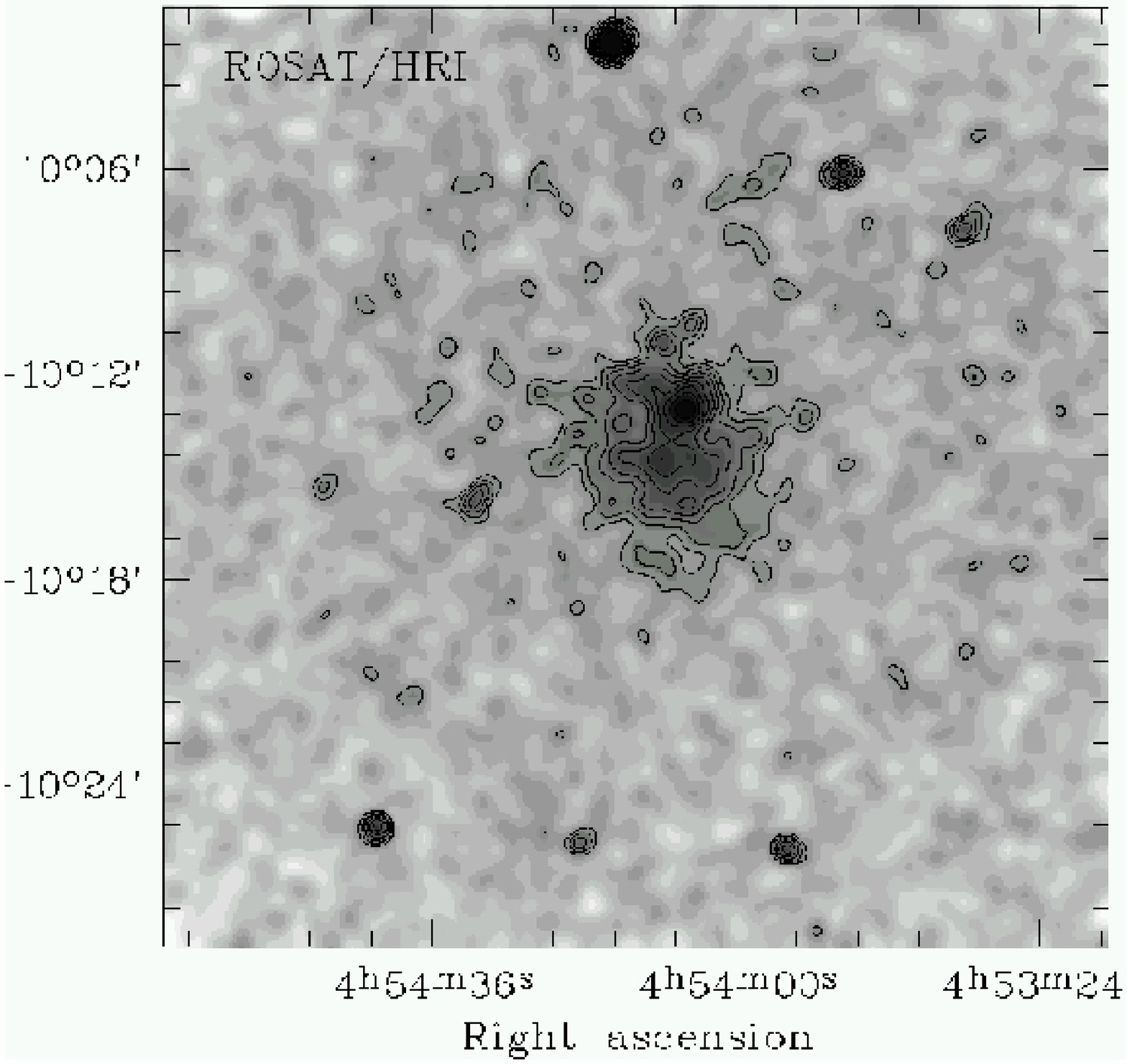,angle=0,height=0.29\textheight}}}
\parbox{0.49\textwidth}{
\centerline{\psfig{figure=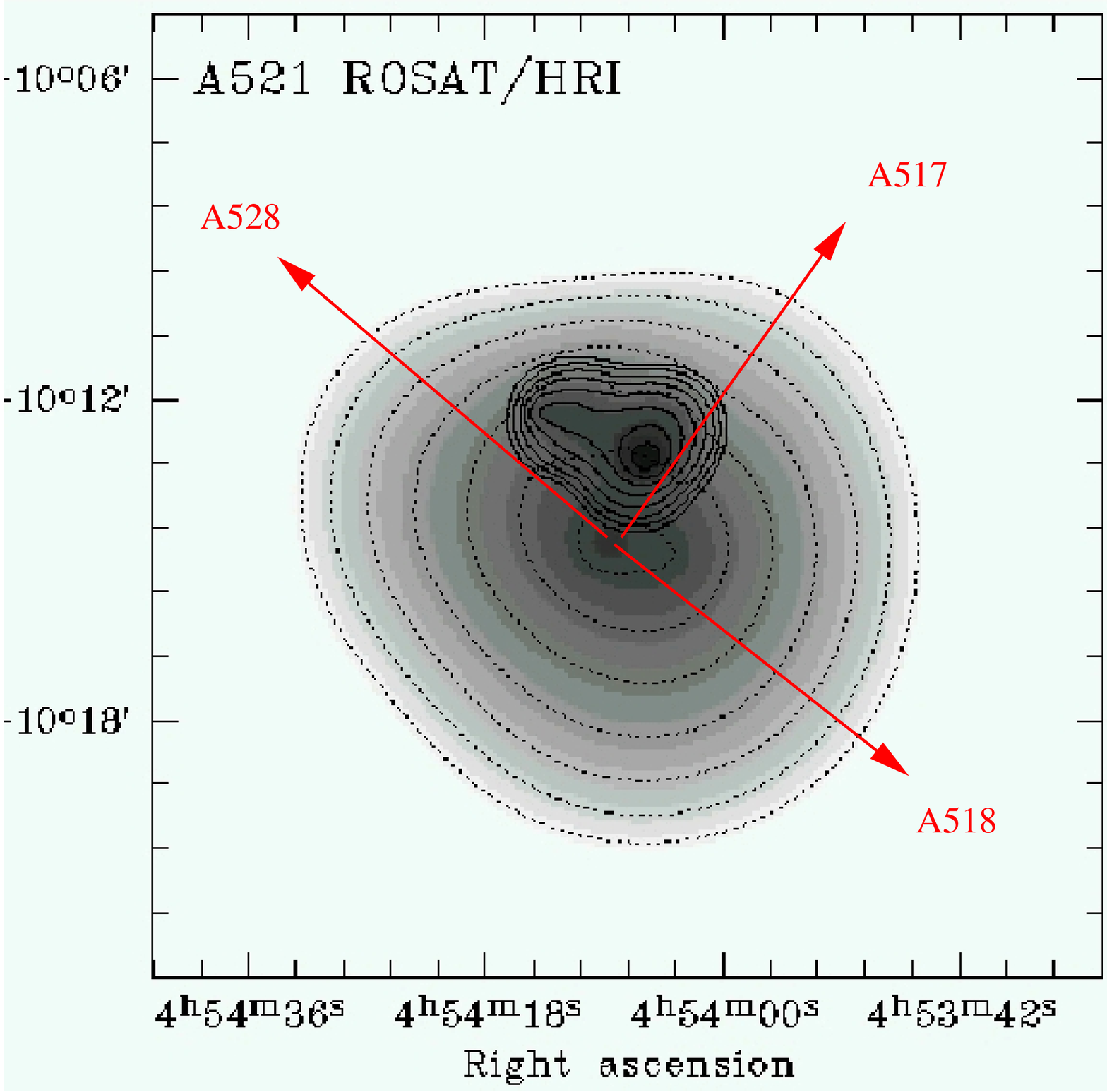,angle=0,height=0.26\textheight}}}   
\caption{\label{fig.wavelet} (Left) {\sl ROSAT} HRI image of A521 
and (Right) the wavelet transformed image (Arnaud et al. 2000).}
\end{figure}   

The power of the wavelet technique is also demonstrated by the
analysis of the {\sl ROSAT} HRI image of A521 by Arnaud et
al. (2000). From visual inspection of the HRI image of A521 one
notices asymmetric isophotes such that the emission peak appears to be
offset from the centers of the fainter isophotes (see left panel in
Figure \ref{fig.wavelet}). Application of the wavelet technique to
this image reveals two distinct structures (see right panel of Figure
\ref{fig.wavelet}). The main cluster appears to be oriented along a
line connecting two adjacent clusters. Nearly perpendicular to this
line is the line connecting the subcluster to the main cluster. This
other line appears to lie nearly parallel to the line pointing to
another adjacent cluster. Consequently, Arnaud et al. conjecture that
A521 lies at the intersection of two large-scale filaments.

\section{Quantitative Classification of Global Morphology}
\label{global}

The other path taken by studies of quantitative X-ray cluster
morphology is to build on the work of Jones \& Forman (1992) and to
devise a quantitative scheme for classifying the morphologies of X-ray
images of galaxy clusters.  As with any classification system in
astronomy the principal motivation for classifying cluster
morphologies is to elucidate fundamental physical properties, in
particular those associated with cluster formation and evolution.

The presence of substructure in clusters implies they are still
forming and evolving dynamically, and thus a logical candidate for a
fundamental parameter is the current dynamical state. The dynamical
state of a cluster is related to the amount of time required for the
cluster to virialize; i.e., a time of order a crossing time. But for a
cluster of a given total mass one can imagine many different
morphological configurations -- and formation histories -- that would
lead to similar relaxation timescales. Hence, to classify clusters
having different formation histories but similar dynamical states we
also require one or more fundamental parameters to specify the type of
merger (e.g., bimodal, many small subclusters) as indicated
qualitatively by the classes of Jones \& Forman (1992).

\subsection{Methods}

Perhaps the most common approach used to quantify the morphologies of
a large number of X-ray cluster images has been with a measure of the
X-ray ellipticity (e.g., McMillan et al. 1989; Davis 1995; Mohr et
al. 1995; Gomez et al. 1997; Gomez, Hughes, \& Birkinshaw 2001;
Kolokotronis et al. 2001). This method is not a particularly good
indicator of the dynamical state since both relaxed and disturbed
clusters can have significant ellipticity. And even disturbed clusters
can have small ellipticity if the substructure is distributed
symmetrically about the cluster center. Moreover, even if both the
ellipticity and associated position angles are considered they only
provide a crude measurement of cluster morphology and have never been
shown to provide an interesting distinction between the variety of
morphologies exemplified by the Jones \& Forman classes.

A better method is the center-shift introduced by Mohr, Fabricant, \&
Geller (1993).  This popular method has been applied in various forms
to X-ray cluster images in several studies (e.g., Mohr et al. 1995;
Gomez et al. 1997, 2000; Rizza et al. 1998; Kolokotronis et al. 2001).
The basic idea is to divide up a cluster image into a series of
circular annuli having different radii but with centers located
initially at a guess for the cluster center. The center-shift is then
given by a weighted average of the centroid computed for each of these
annuli.

Since the center-shift is sensitive only to asymmetries in the X-ray
images (in particular non-ellipsoidal configurations) it is a much
more reliable than the ellipticity as an indicator for when a cluster
is relaxed. However, it is not transparent how the center shift
translates into a physical measure of the dynamical state.  And since
the center-shift is most sensitive to mergers of equal-mass
subclusters, it cannot by itself distinguish the full range of
structures exhibited by the Jones \& Forman morphological classes.

If the only objective were to distinguish the full range of cluster
morphologies then the logical procedure would be to decompose cluster
images into a set of orthogonal basis functions of which wavelets (see
\S \ref{detailed}) are the probably best example. The wavelet coefficients
would then define the parameter space of cluster
morphologies. Unfortunately, there is no obvious connection (of which
I am aware) between wavelet coefficients and a physical measure of the
dynamical state.

One method that is both closely related to the cluster dynamical state
and provides a quantitative description of the full range of Jones \&
Forman morphological classes is the ``power ratio'' method (Buote \&
Tsai 1995, 1996; Buote 1998). The power ratios are constructed from
the moments of the two-dimensional gravitational
potential. Specifically, one evaluates the square of the moments over
a circle of radius, $R$, where the origin is located at the center of
mass or the at the largest mass peak. The ratio of term, $m$, to the
monopole term is called a ``power ratio'',
\begin{equation}
{P_m\over P_0} \equiv
{ \langle(\Psi^{\rm int}_m)^2\rangle\over \langle(\Psi^{\rm int}_0)^2\rangle},
\label{eqn.prs}
\end{equation}
where $\Psi_m^{\rm int}$ is the $m$th multipole of the two-dimensional
gravitational potential due to matter interior to the circle of
radius, $R$, and $\langle\cdots\rangle$ represents the azimuthal
average around the circle. In detail we have,
\begin{equation}
P_0=\left[a_0\ln\left(R\right)\right]^2,\label{eqn.power0}
\end{equation}
for $m=0$,
\begin{equation}
P_m={1\over 2m^2 R^{2m}}\left( a^2_m + b^2_m\right)\label{eqn.powerm}
\end{equation}
for $m>0$. The moments $a_m$ and $b_m$ are given by,
\begin{eqnarray}
a_m(R) & = & \int_{R^{\prime}\le R} \Sigma(\vec x^{\prime})
\left(R^{\prime}\right)^m \cos m\phi^{\prime} d^2x^{\prime}, \nonumber \\
b_m(R) & = & \int_{R^{\prime}\le R} \Sigma(\vec x^{\prime})
\left(R^{\prime}\right)^m \sin m\phi^{\prime} d^2x^{\prime}, \nonumber
\end{eqnarray}
where $\vec x^{\prime} = (R^{\prime},\phi^{\prime})$.

These ratios are directly related to the 2D gravitational potential if
one has a map of the 2D surface mass density such as provided by weak
gravitational lensing studies. For X-ray studies $\Sigma$ is replaced
with the X-ray surface brightness, $\Sigma_{\rm x}$, and therefore the
power ratios in X-ray studies are really derived from a pseudo
potential. These ratios are most sensitive to structures on the same
scale as the aperture radius, $R$.

When the aperture is located at the peak of the X-ray emission the
dipole power ratio, $P_1/P_0$, provides structural information similar
to the center shift discussed above (see also Dutta 1995). For an
aperture located at the centroid of the surface brightness the dipole
moment vanishes. In this case the quadrupole power ratio, $P_2/P_0$,
is sensitive to the degree of flattening and is related to the
ellipticity. But unlike ellipticity $P_2/P_0$ is also sensitive to the
radial profile of the X-ray emission.

The primary physical motivation behind the power ratios is that they
are related to potential fluctuations. And since it is thought that
large potential fluctuations drive violent relaxation in clusters, the
power ratios are closely related to the dynamical state of a cluster
(Buote 1998). The other motivation is that the multipoles are a
complete orthogonal set of basis functions for the (pseudo) potential
and thus are well suited to classify the wide range of observed
cluster morphologies.

To get a feel for the power ratios let us see how they behave on the
{\sl ROSAT} PSPC images of clusters in the different Jones \& Forman
morphological classes shown in Figure \ref{fig.refclusters}. The four
clusters inhabit the extreme Jones \& Forman classes. A2029 is a
smooth, single component system apparently in a relaxed state. A85 has
a regular dominant component but with a small structure $\sim 0.6$ Mpc
to the S. A1750 is a double cluster consisting of two roughly
equal-sized components separated by $\sim 1$ Mpc. A514 is a highly
irregular aggregation of structures.

\begin{figure}[th]
\centerline{\psfig{figure=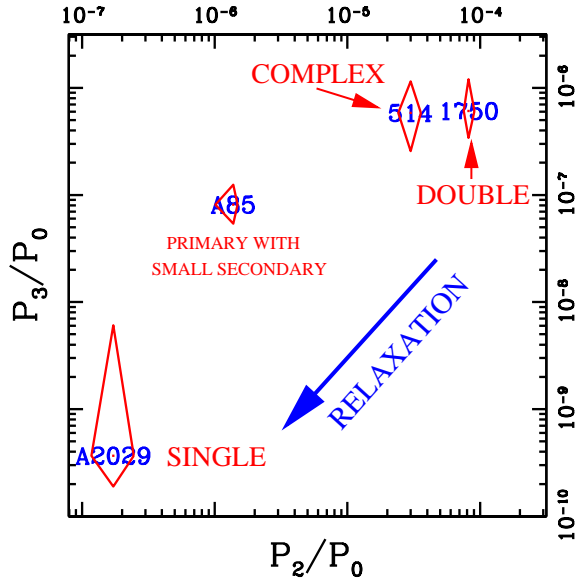,angle=0,height=0.4\textheight}}
\caption{\label{fig.jf_1mpc} Power ratios (from Buote \& Tsai 1996)
for the clusters in Figure \ref{fig.refclusters} computed within a
circular aperture of 1 Mpc radius located at the centroid of the X-ray
emission.}
\end{figure} 

In Figure \ref{fig.jf_1mpc} I show the power ratios, $P_2/P_0$ and
$P_3/P_0$, of these clusters computed for a 1 Mpc aperture\footnote{In
Buote \& Tsai (1995, 1996) $H_0=80$ km s$^{-1}$ Mpc$^{-1}$ and $q_0=0$
was assumed.} where the aperture is located at the centroid of the
X-ray emission (i.e., analog of the center of mass).  It can be seen
that the single-component cluster is well separated from the primary
with small secondary. And each of these classes is clearly
distinguished from the disturbed complex and double clusters. In
effect the power ratios have defined a morphological evolutionary
track where the young, unrelaxed clusters are born at the top right of
the figure. As they relax and erase their substructure they pass
through a phase similar to A85 until they are old and evolved systems
like A2029.

\begin{figure}[ht]
\centerline{\psfig{figure=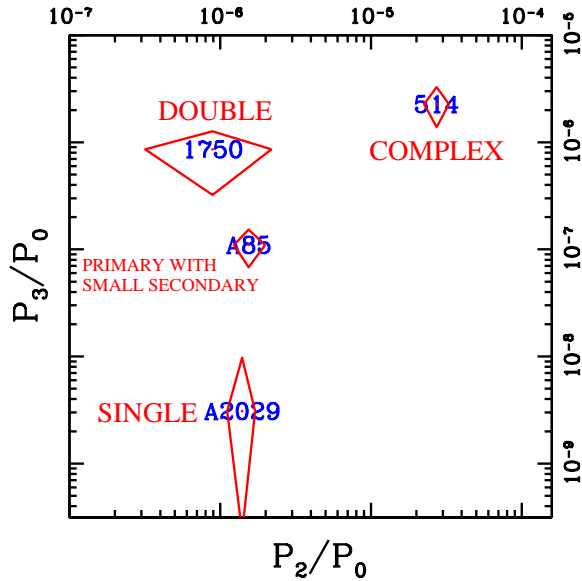,angle=0,height=0.4\textheight}}
\caption{\label{fig.jf_05mpc} As Figure \ref{fig.jf_1mpc} but for the
0.5 Mpc aperture.}
\end{figure}

Although we have succeeded in obtaining a successful broad
classification according to dynamical states, we have not
distinguished clearly between the different classes of highly
disturbed clusters (i.e., complex and double). Since there is nothing
special about the 1 Mpc aperture it is sensible to explore the effects
of using different apertures. The result of computing the power ratios
in a 0.5 Mpc aperture are displayed in Figure \ref{fig.jf_05mpc}

By focusing initially on $P_2/P_0$ it can be seen that three of the
clusters appear to be relaxed systems (i.e., small $P_2/P_0$). This is
because the 0.5 Mpc aperture only encloses 1 component of the double
cluster and only the primary component of A85. The single component
cluster A2029 appears relaxed on both the 0.5 and 1 Mpc
scales. However, A514 is complex on many scales and it is easily
distinguished from the other reference clusters as a disturbed system
in the 0.5 Mpc aperture. Of course, one only needs to appeal to
$P_3/P_0$ to verify that both the double and complex clusters are
actually in a younger dynamical state than the others. {\bf Hence, the
power ratios represent a quantitative implementation of the Jones \&
Forman classification scheme, particularly on the 0.5 Mpc scale.}

\subsection{Merger Frequency of ROSAT Clusters}

\begin{figure}[ht]
\parbox{0.49\textwidth}{
\centerline{\psfig{figure=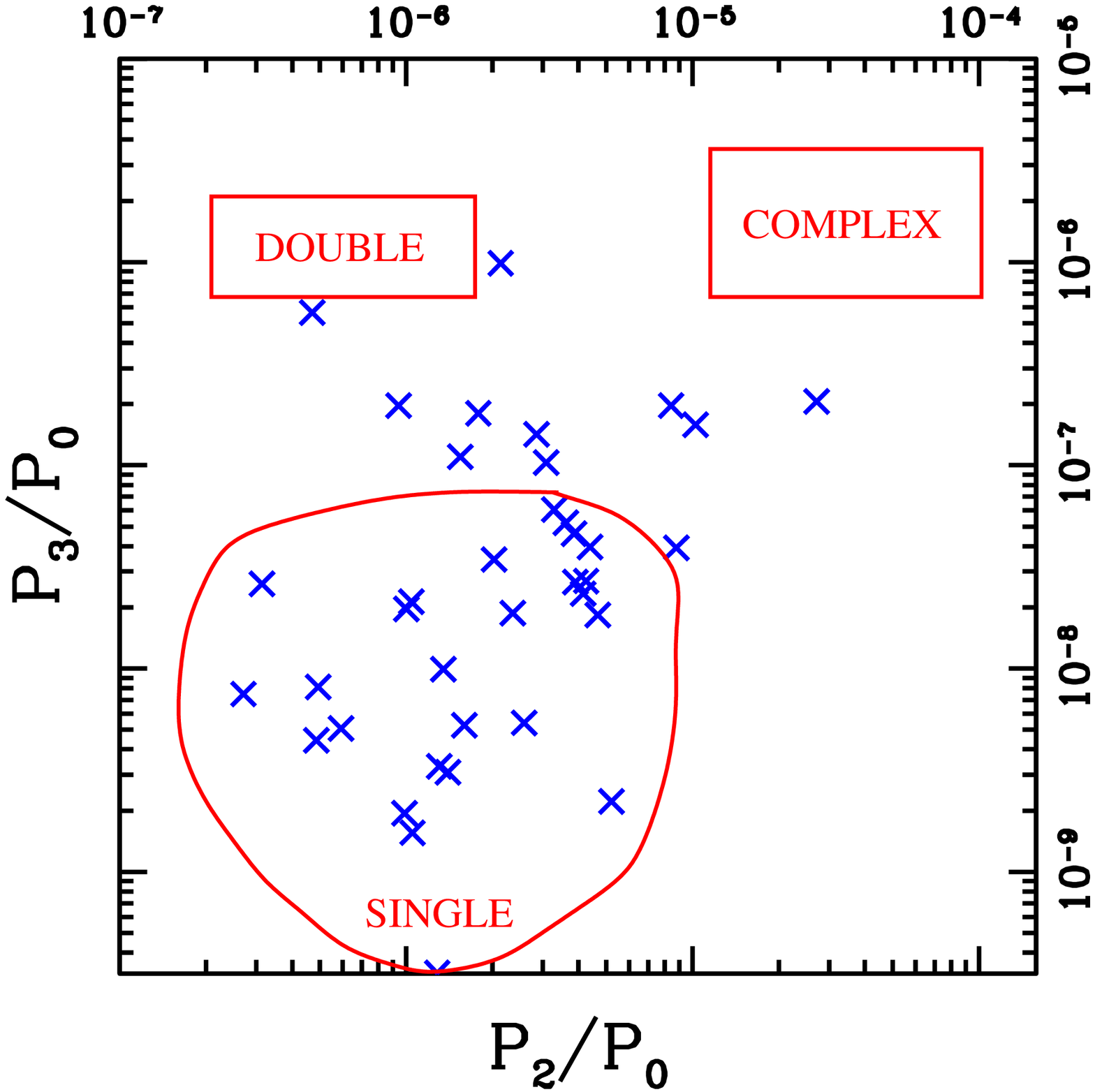,angle=0,height=0.315\textheight}}}
\parbox{0.49\textwidth}{
\centerline{\psfig{figure=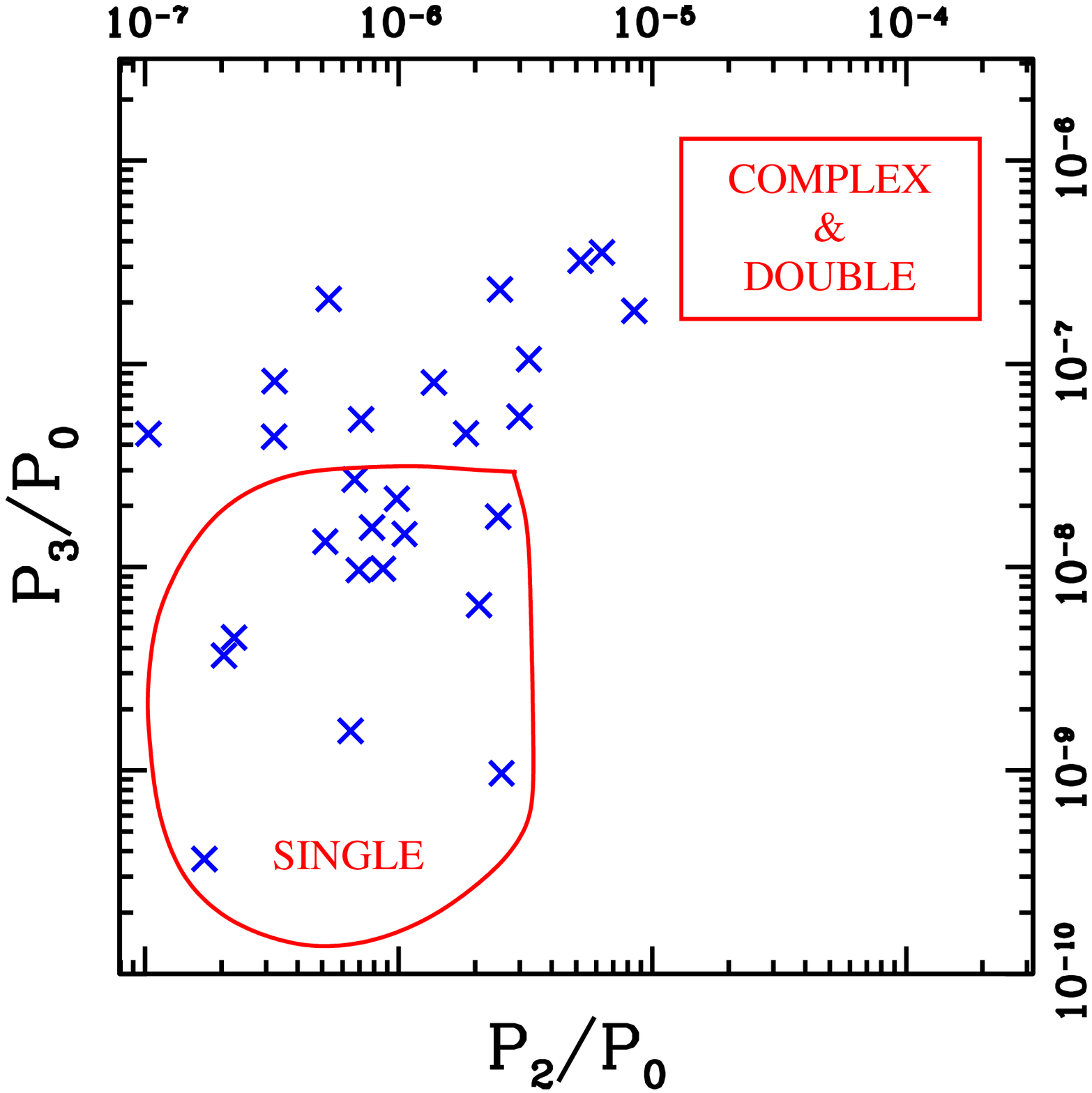,angle=0,height=0.315\textheight}}} 
\caption{\label{fig.pr_rosat} Power ratios of the brightest $\sim 40$
clusters (Buote \& Tsai 1996) computed within apertures of 0.5 Mpc
(Left) and 1 Mpc (Right).}
\end{figure} 

The result of computing power ratios for the brightest $\sim 40$ {\sl
ROSAT} clusters is displayed in Figure \ref{fig.pr_rosat}. It is
immediately apparent that there is a marked deficiency of highly
disturbed clusters (complex and double). These brightest clusters
therefore lack young members and are instead dominated by mostly
evolved clusters with only small-scale ($<500$ kpc) substructure.
Since such highly evolved clusters are usually associated with cooling
flows it should be expected that cooling flows dominate the brightest
clusters as has been suggested before on different grounds (e.g.,
Arnaud 1988; Forman \& Jones 1990; Edge et al. 1992; Peres et
al. 1998).

\begin{figure}[ht]
\centerline{\psfig{figure=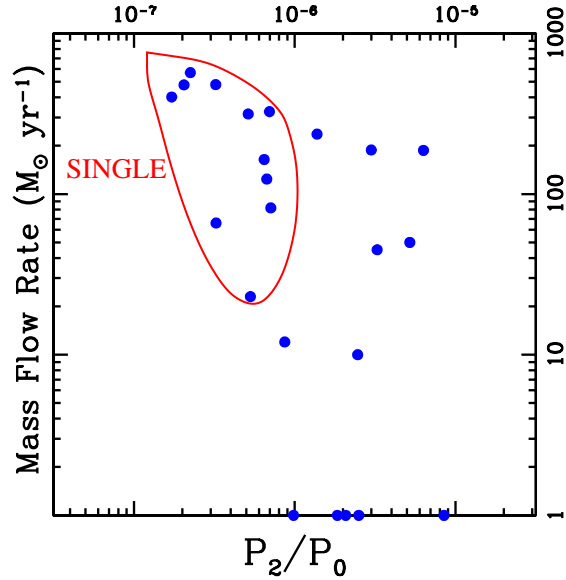,angle=0,height=0.4\textheight}}   
\caption{\label{fig.cf} As Figure \ref{fig.pr_rosat} for the 1 Mpc
aperture except that the cooling flow mass deposition rate has been
plotted on the vertical axis.}
\end{figure} 

In Figure \ref{fig.cf} the quantitative connection between cooling flows
and cluster morphology is shown by the anti-correlation of the mass
deposition rate ($\dot{M}$) and $P_2/P_0$. This represents the first
quantitative description of the anti-correlation of substructure with
the strength of a cooling flow. Note the large scatter for systems
that have significant substructure (i.e., large $P_2/P_0$).  Analysis
of this correlation and its large scatter should shed light on how
cooling flows are disrupted by merges and are subsequently
re-established.

\section{High-Redshift Clusters}

\begin{figure}[ht]
\parbox{0.49\textwidth}{
\centerline{\psfig{figure=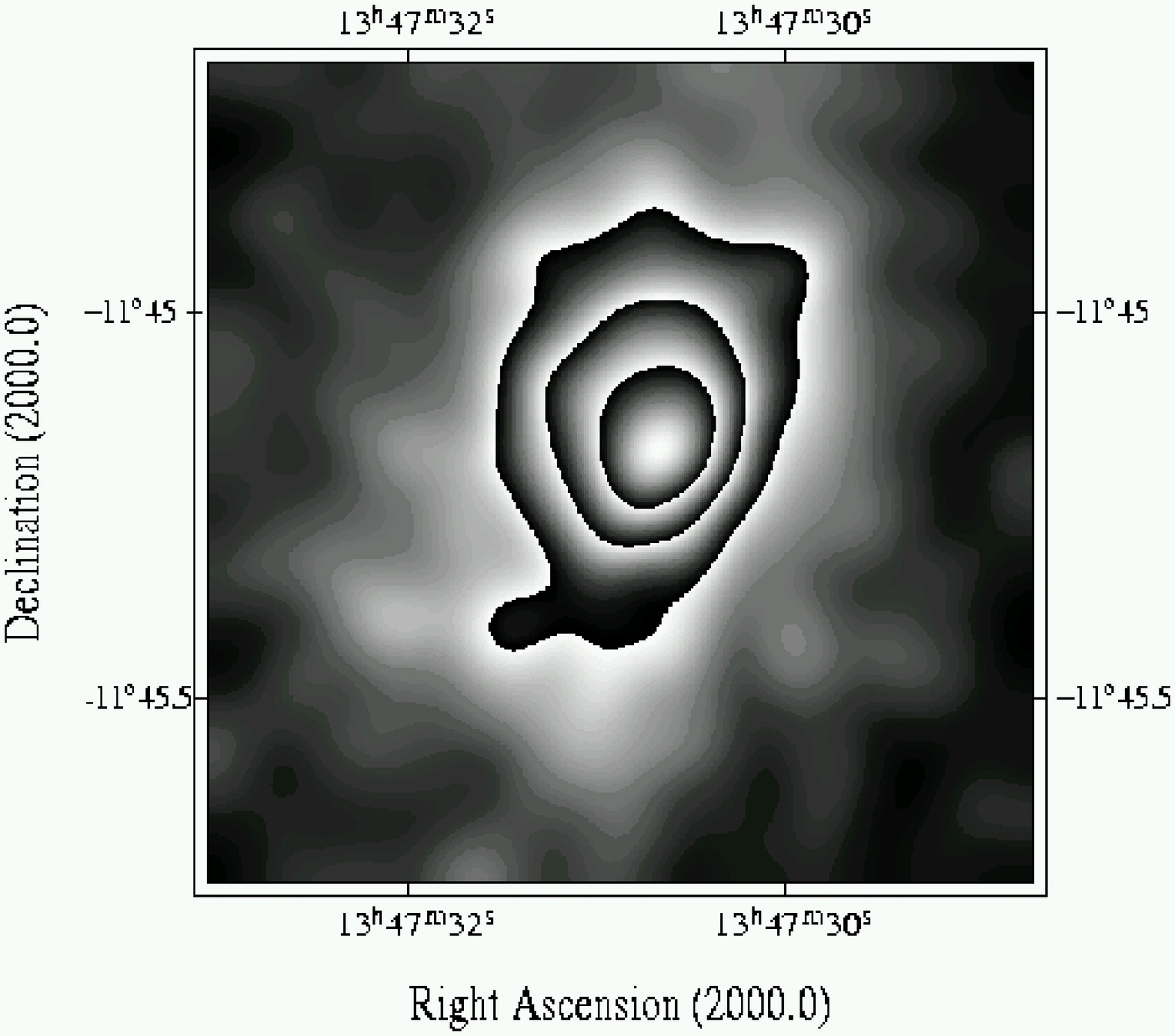,angle=0,height=0.315\textheight}}}
\parbox{0.49\textwidth}{
\centerline{\psfig{figure=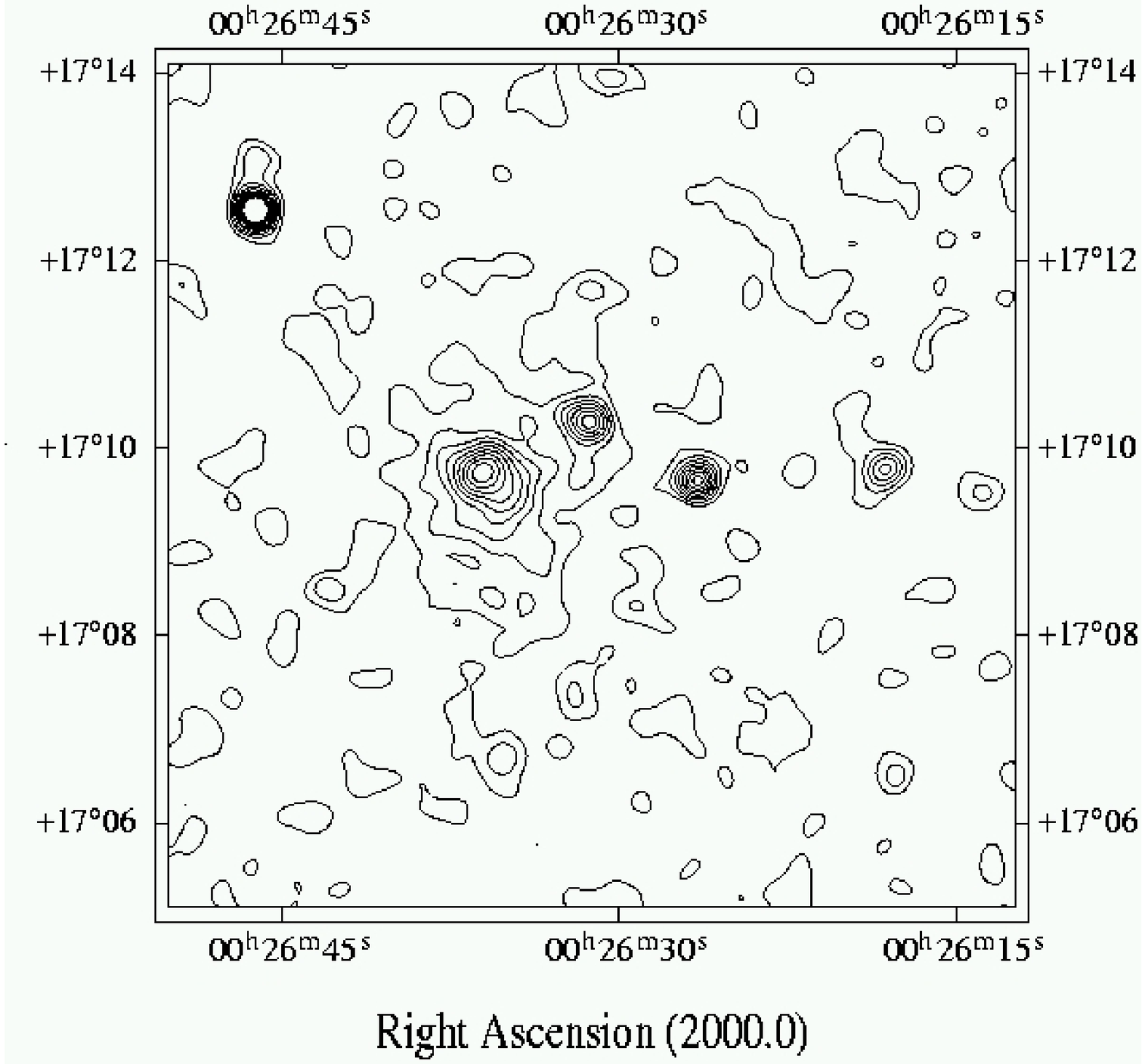,angle=0,height=0.285\textheight}}}
\caption{\label{fig.highz} {\sl ROSAT} HRI images of (Left)
RXJ1347-1145 from Schindler et al. (1997) and (Right) CL0024+17 from
B\"{o}hringer et al. (1997).}
\end{figure} 
     
Unfortunately, because of the limited resolution and collecting area
of {\sl ROSAT} it has been difficult to study the morphologies of
distant clusters. Two of the best examples ($z\sim 0.4$) imaged with
the {\sl ROSAT} HRI are displayed in Figure \ref{fig.highz}. The
cluster RXJ1347.5-1145 appears to be a relaxed, cooling flow
(Schindler et al. 1997) while the cluster Cl0024+17 may have
substantial substructure as quantified by a center shift
(B\"{o}hringer et al. 2000). These tantalizing glimpses demonstrate
the need for a systematic study at high resolution with {\sl Chandra}.

\section{Morphology and Cosmology}

Fossil imprints of the process of formation are retained in the
cluster substructure. In the standard hierarchical paradigm of
structure formation the mass spectrum of subclusters is related to the
power spectrum of mass density fluctuations which is a key
distinguishing property of cosmological models (e.g., Peacock
1999). As clusters evolve dynamically the mass spectrum of subclusters
changes. In a standard Friedmann-Robertson-Walker universe with
$\Omega_0<1$ and $\lambda_0=0$, the linear growth of density
fluctuations becomes strongly suppressed when the curvature term in
the Friedmann equation exceeds the matter term.  The redshift
delineating this transition from an Einstein - de Sitter phase to one
of free expansion is then $1+z_{\rm trans}=\Omega_0^{-1}-1$; i.e. when
the matter density $\Omega(z_{\rm trans})=0.5$. Hence, if $\Omega_0\ll
1$, then objects formed a long time in the past relative to universes
where $\Omega_0\approx 1$, and thus clusters in low-density universes
should be, on average, more relaxed than clusters in universes with
$\Omega_0\approx 1$.

\subsection{Semi-Analytical Models}

\begin{figure}[ht]
\centerline{\psfig{figure=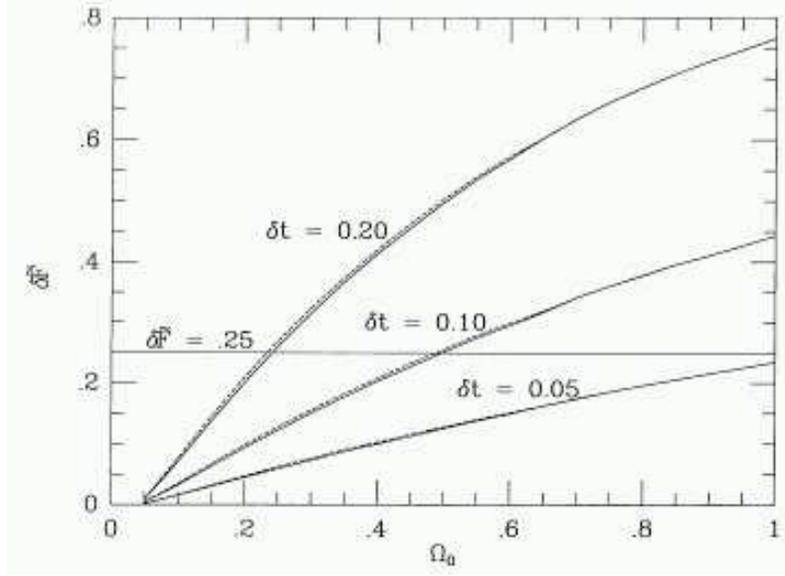,angle=0,height=0.4\textheight}}   
\caption{\label{fig.rlt} Figure 3 from Richstone et
al. (1992). Fraction of present-day clusters which formed within the
last time interval, $\delta t$, as a function of $\Omega_0$. The time
intervals are in units of $H_0^{-1}$. The horizontal line represents
a conservative observational estimate of the frequency of substructure.}
\end{figure} 

Richstone, Loeb, \& Turner (1992) presented the first theoretical
model relating $\Omega_0$ to the observed frequency of substructure in
clusters. In their semi-analytical calculations they avoided the issue
of the power spectrum by concentrating on clusters having the same
total mass.  The collapse time of a $10^{15}h^{-1}M_{\sun}$ spherical
density perturbation (taken to be twice the turn around time) was
defined to be the dividing point between clusters that do and do not
possess substructure. By further assuming that any substructure is
erased on a crossing time (taken to be $0.1/H_0$), Richstone et
al. computed the quantity $\delta F$, the fraction of present-day
clusters which formed within the last time interval, $0.1/H_0$, as a
function of $\Omega_0$ and $\lambda_0$. They found $\delta
F\sim\Omega_0$ (see Figure \ref{fig.rlt}). When compared to the
estimates of $\gtrsim 30\%$ for the frequency of substructure in
nearby clusters (Jones \& Forman 1992) Richstone et al. concluded that
$\Omega_0\gtrsim 0.5$.

Follow up theoretical studies by Kauffmann \& White (1993), Lacey \&
Cole (1993), and Nakamura, Hattori, \& Mineshige (1995) emphasized
that the time for substructure to be erased is variable and can be
especially long for substructures with compact cores. The relationship
between the collapse time of a spherical density perturbation and
subclustering, though qualitatively reasonable, is ambiguous.
Consequently, it is difficult to compare directly the frequency of
observed substructure to predictions of semi-analytic models based on
Richstone et al.'s idea.

Thus, a fundamental limitation of these studies is that they only
predict the ambiguous ``frequency of substructure'' rather than a
well-defined quantitative measure of cluster morphology such as the
power ratios. Since Richstone et al.'s idea is really a statement
about the dynamical states of clusters, in Buote (1998) I used a
related (but more detailed) semi-analytical approach to study the
behavior of cluster power ratios in different cosmologies. Violent
relaxation (Lynden-Bell 1967) is the key process driving the
elimination of large potential fluctuations.  It operates on a
timescale of $\sim 1-2$ crossing times and proceeds independently of
the masses of the constituents. Consequently, I argued that a
plausible definition of the dynamical state of a cluster is,
\begin{equation}
{\langle(\Delta\Phi^{\rm
int})^2\rangle\over\langle(\overline{\Phi}^{\rm int})^2
\rangle}
\approx \left({\Delta M \over \overline{M}}\right)^2 +
\sum_{l>0} {\langle({\Phi^{\rm int}_l})^2\rangle \over
\langle(\Phi^{\rm int}_0)^2\rangle},
\label{eqn.key}
\end{equation}
where $\overline{M}$ is the average mass and $\Delta M$ is the mass
accreted over a relaxation time, typically assumed to be a crossing
time; ${\Delta M / \overline{M}}$ is called the ``fractional accreted
mass''. This equation states that over the duration of a crossing time
the fractional increase in the rms spherically averaged potential is
approximately equal to the fractional increase in the mass added in
quadrature to the ratios of the increases of the rms spherically
averaged higher order potential multipoles to the monopole.

The key premise is that the amount of accreted mass over the previous
relaxation timescale determines the amount of substructure (or
non-ellipsoidal distortions) which is similar to the premise of
Richstone et al. that substructure is related to the collapse and
crossing times. This premise requires that $\Delta M / \overline{M}$
be strongly correlated with the other low-order terms, which are
approximately,
$\langle(\Phi^{int}_l)^2\rangle/\langle(\Phi^{int}_0)^2\rangle$,
defined at the epoch of interest. These terms are just the 3D versions
of the power ratios (see equation \ref{eqn.prs}).

\begin{figure}[ht]
\parbox{0.49\textwidth}{
\centerline{\psfig{figure=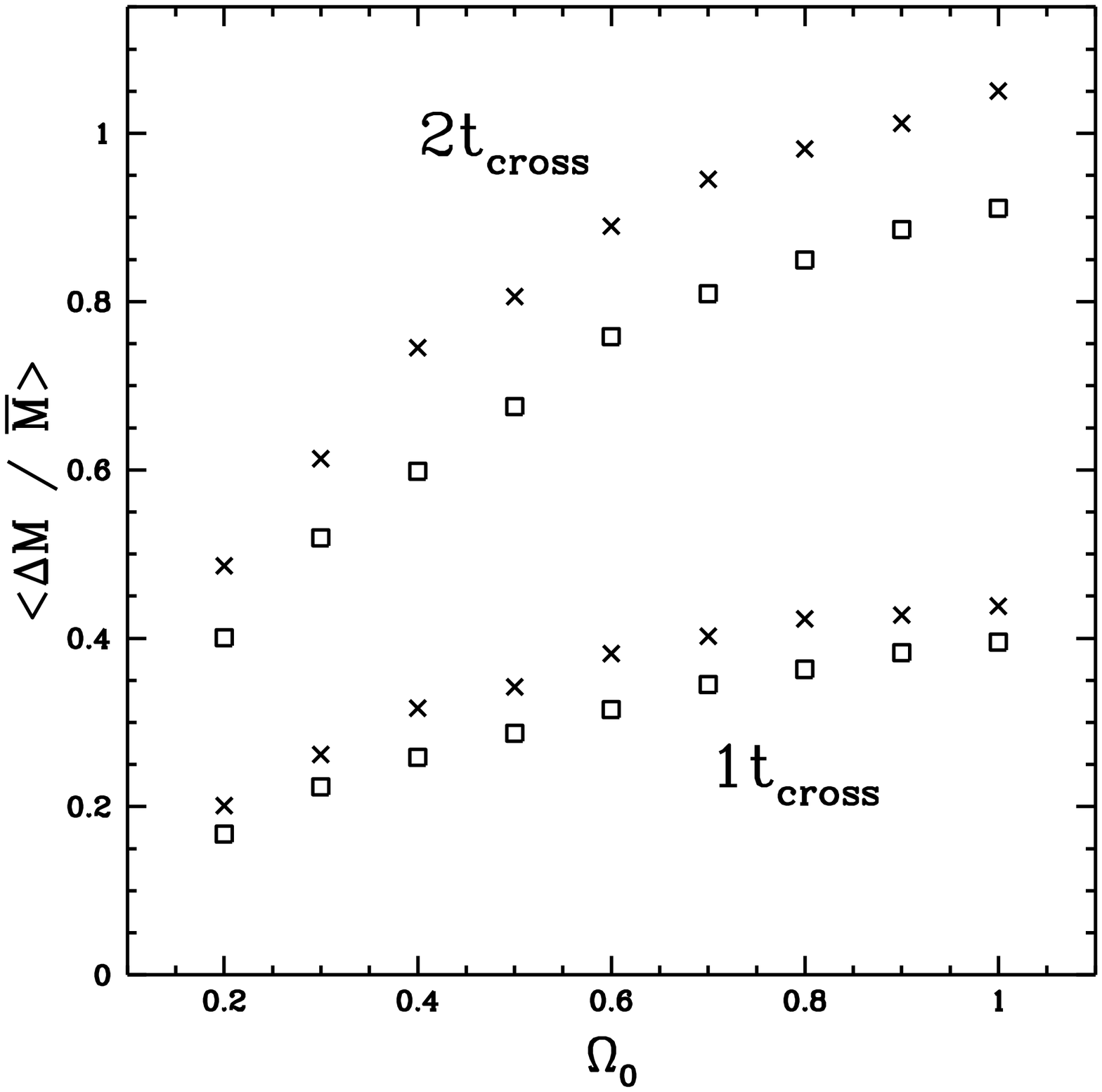,angle=0,height=0.325\textheight}}}
\parbox{0.49\textwidth}{
\centerline{\psfig{figure=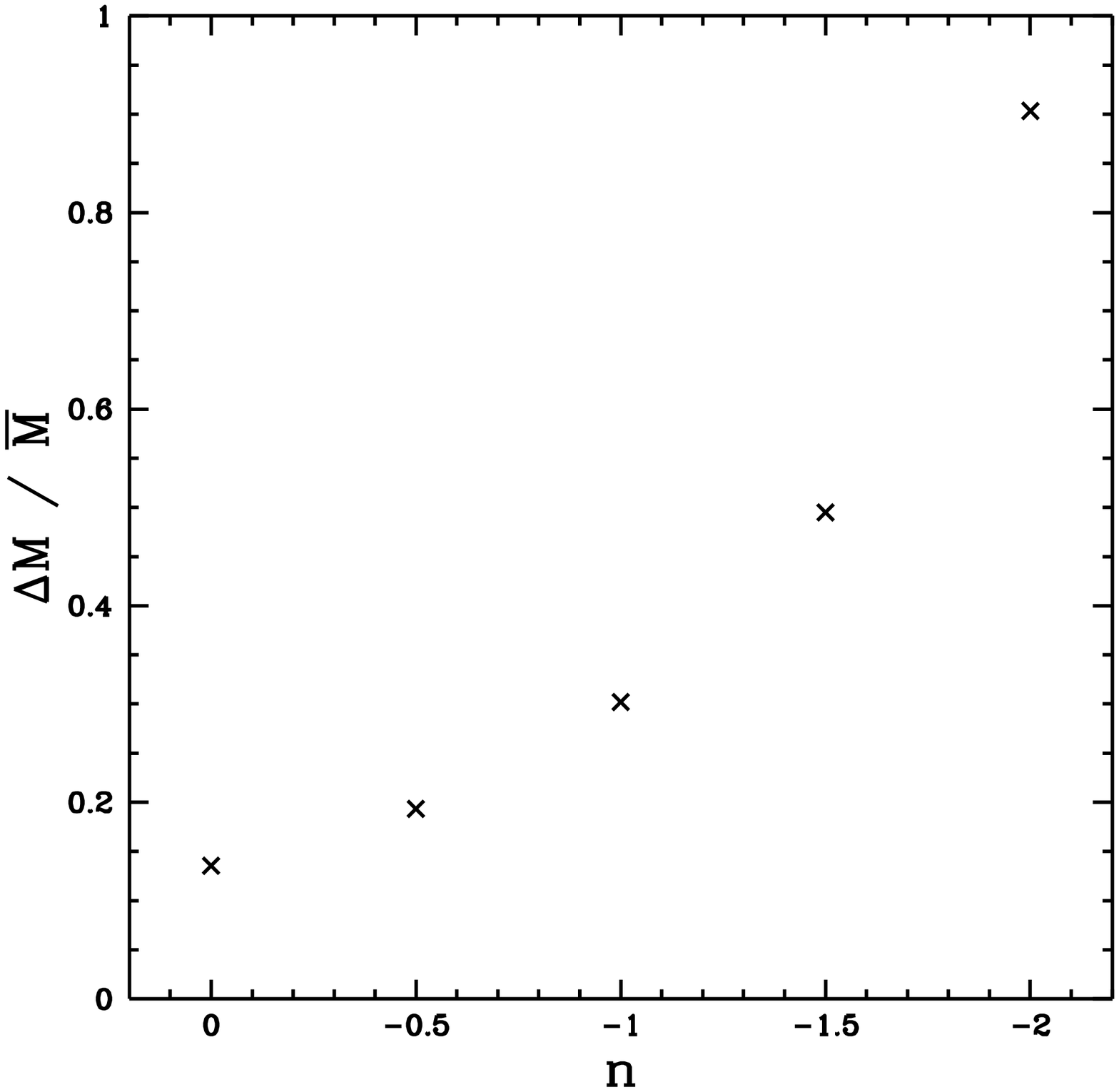,angle=0,height=0.325\textheight}}}
\caption{\label{fig.b98} From Buote (1998): (Left) The mass-averaged
fractional accreted mass evaluated for $r=1h^{-1}$ Mpc at $z=0$ for
CDM models as a function of $\Omega_0$ $(\lambda_0=0)$. The crosses
indicate a mass average over the full range $(0.35-3)\times
10^{15}h^{-1}M_{\sun}$, and the boxes indicate a lower limit of
$7\times 10^{14}h^{-1}M_{\sun}$.  Relaxation timescales of 1 and 2
crossing times are shown. (Right) $\Delta M /\,\overline{M}$ for
clusters of mass $7\times 10^{14}h^{-1}M_{\sun}$ evaluated for
$r=1h^{-1}$ Mpc at $z=0$ for models with $\Omega_0=1$ and power
spectra $P(k)\propto k^n$ as a function of spectral index $n$.}
\end{figure}

The dependence of $\Delta M /\,\overline{M}$ on $\Omega_0$ and the
power spectrum is shown in Figure \ref{fig.b98}. We see the expected
increase in fractional accreted mass with increasing $\Omega_0$ where
$\Delta M /\,\overline{M}\propto \sqrt{\Omega_0}$, but the
normalization does depend sensitively on the assumed relaxation
timescale similarly to the previous related studies by Richstone et
al. and others. Also shown is the dependence on $n$, the spectral
index of models with $P(k)\propto k^n$, which is considerably steeper,
$\Delta M /\,\overline{M}\propto (-n)^{2.5}$. Since the observable
low-order power ratios should behave as $P_m/P_0\sim (\Delta M
/\,\overline{M})^2$ (see section 2.2 of Buote 1998), the power-ratio
distribution for a large sample of clusters should be an interesting
probe of $\Omega_0$ and the power spectrum.

\subsection{N-Body Simulations}

N-body simulations of CDM clusters confirm that the mean value of
$P_m/P_0$ for small $m$ in a cluster sample increases with $\Omega_0$
(Buote \& Xu 1997; Thomas et al. 1998). But Buote \& Xu (1997) also
perform simulations with $P(k)\propto k^n$ for different $n$ and find
that the mean value of $P_m/P_0$ is barely affected by $n$. On the
other hand they find that $n$ does affect significantly the variance
of $P_m/P_0$.  ($\Omega_0$ does not seem to affect the variance.) 
These conclusions have to be viewed with some caution because these
dark-matter-only simulations analyze the projected square of the mass
density in an attempt to mimic X-ray observations. Further work with
large high-resolution N-body simulations is required to establish
precisely the relationships between $\Delta M /\,\overline{M}$,
$P_m/P_0$, $\Omega_0$, and $P(k)$ (and $\lambda_0$).

Other N-body simulations with and without gas show that center-of-mass
shifts are also sensitive to $\Omega_0$ (Jing et al. 1995; Crone,
Evrard, \& Richstone 1996). Generally both semi-analytic models and
dark-matter-only N-body simulations agree that center shifts and power
ratios can distinguish between CDM models with different values of
$\Omega_0$.  The same holds for gas-dynamical N-body simulations
(Evrard et al. 1993; Mohr et al. 1995)

\begin{figure}[ht]
\parbox{0.49\textwidth}{
\centerline{\psfig{figure=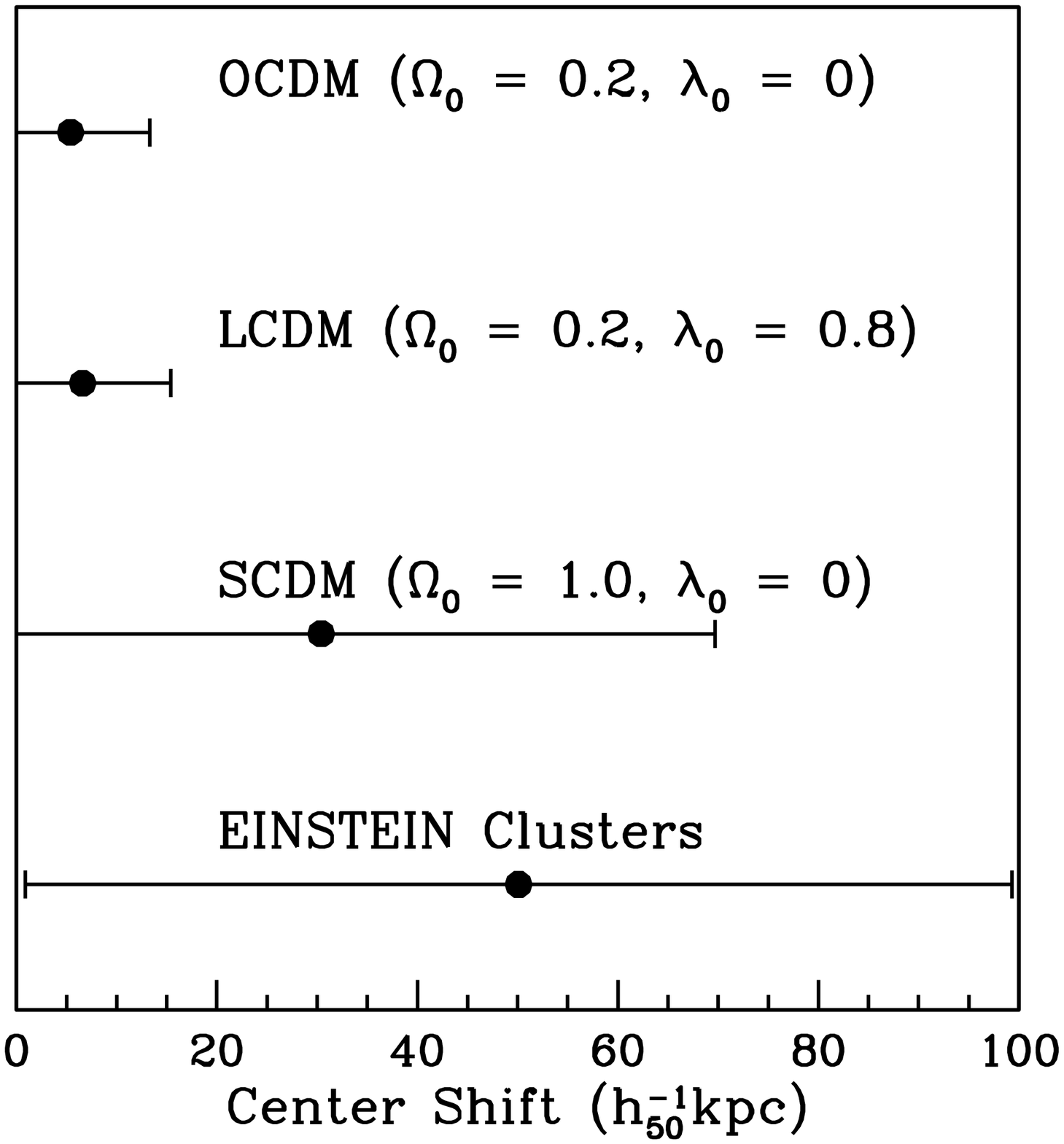,angle=0,height=0.325\textheight}}}
\parbox{0.49\textwidth}{
\centerline{\psfig{figure=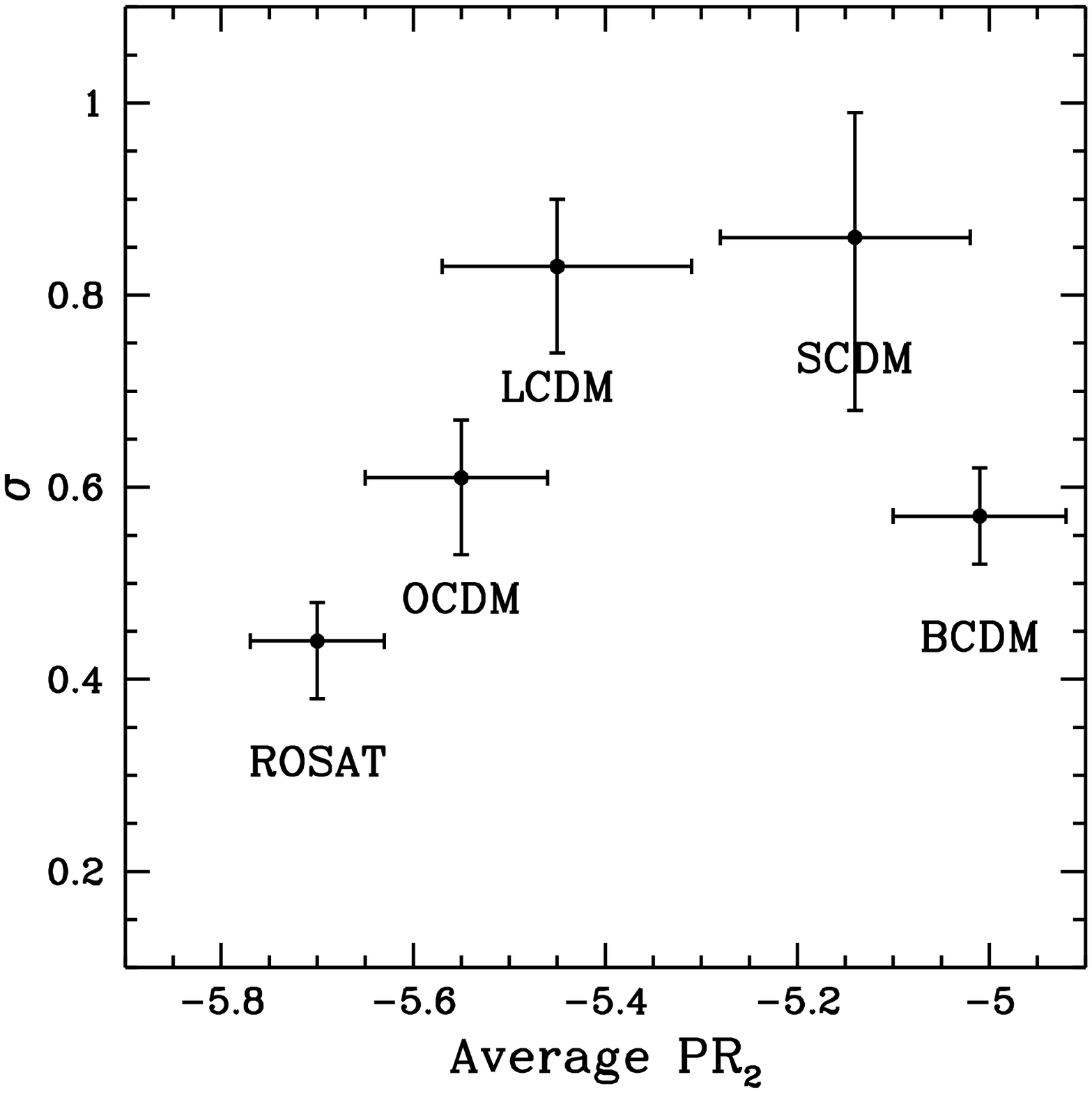,angle=0,height=0.325\textheight}}}
\caption{\label{fig.mohr_bx} (Left) Center shifts obtained by Mohr et
al. (1995) from simulated and {\sl Einstein} clusters. (Right) Power
ratios ($PR_2=\log_{10}(P_2/P_0)$) obtained by Buote \& Xu (1997) for
{\sl ROSAT} and simulated clusters: OCDM ($\Omega_0=0.35$), LCDM
($\Omega_0=0.35, \lambda_0=0.65$), and SCDM ($\Omega_0=1.0$).}
\end{figure}  

However, when the N-body simulations (with or without gas) are
compared to X-ray observations of clusters conflicting results are
obtained (Figure \ref{fig.mohr_bx}). Mohr et al. (1995) compare center
shifts of clusters formed in hydrodynamical simulations to {\sl
Einstein} clusters and conclude that $\Omega_0\approx 1$ whereas Buote
\& Xu (1997) compare power ratios of the projected square of the dark matter
density to {\sl ROSAT} clusters and conclude
$\Omega_0<1$. Furthermore, the clusters formed in the hydrodynamical
simulations by Valdarnini, Ghizzardi, \& Bonometto (1999) give power
ratios different from those obtained by Buote \& Xu (1997).

All of these simulations have deficiencies. The most important
deficiency in the hydrodynamical simulations is the poor force
resolution for the gas: softening lengths of $\sim 80$ kpc for
Valdarnini et al. (1999) and over 100$h^{-1}$ kpc for Mohr et
al. (1995) . The simulations of Mohr et al. also contained only six
clusters which is too small for statistical studies. Finally, the
simulations of Buote \& Xu (1997) approximated the gas distribution
using the dark matter.

Clearly until appropriate simulations are applied to this problem we
will not have a reliable constraint on $\Omega_0$ or $P(k)$ from
cluster morphologies. What is needed are high-resolution ($\la 20$
kpc) three-dimensional gas-dynamical simulations of a large number
($\ga 50$) of clusters. The existing observational samples of {\sl
Einstein} data (Mohr et al. 1995) and {\sl ROSAT} data (Buote \& Tsai
1996) also need to be expanded and re-analyzed with new
high-resolution, high S/N {\sl Chandra} and {\sl XMM} data. These
requirements are not excessive for a problem that deserves serious
attention.

\section{Morphology and Radio Halos}

It has been noticed for some time that X-ray observations provide
circumstantial evidence for a connection between cluster merging and
radio halos (see Feretti 2000 and references therein) because, in
particular, radio halos are only found in clusters possessing X-ray
substructure and weak (or non-existent) cooling flows. However, it has
been argued (e.g., Giovannini \& Feretti 2000; Liang et al. 2000;
Feretti 2000) that merging cannot be solely responsible for the
formation of radio halos because at least 50\% of clusters show
evidence for X-ray substructure (Jones \& Forman 1999) whereas only
$\sim 10\%$ possess radio halos. (Note X-ray and optical substructures
are well-correlated -- Kolokotronis et al. 2001.)

Unfortunately, it is difficult to interpret the importance of merging
using the observed frequency of substructure as it does not itself
quantify the deviation of an individual cluster from a virialized
state. And the shocks that could be responsible for particle
acceleration will be proportionally stronger in clusters (of the same
mass) with the largest departures from a virialized state.  To measure
the dynamical states of clusters from X-ray images it is necessary to
quantify the cluster morphologies using statistics such as the
center-shift and the power ratios.

\begin{figure}[ht]
\parbox{0.49\textwidth}{
\centerline{\psfig{figure=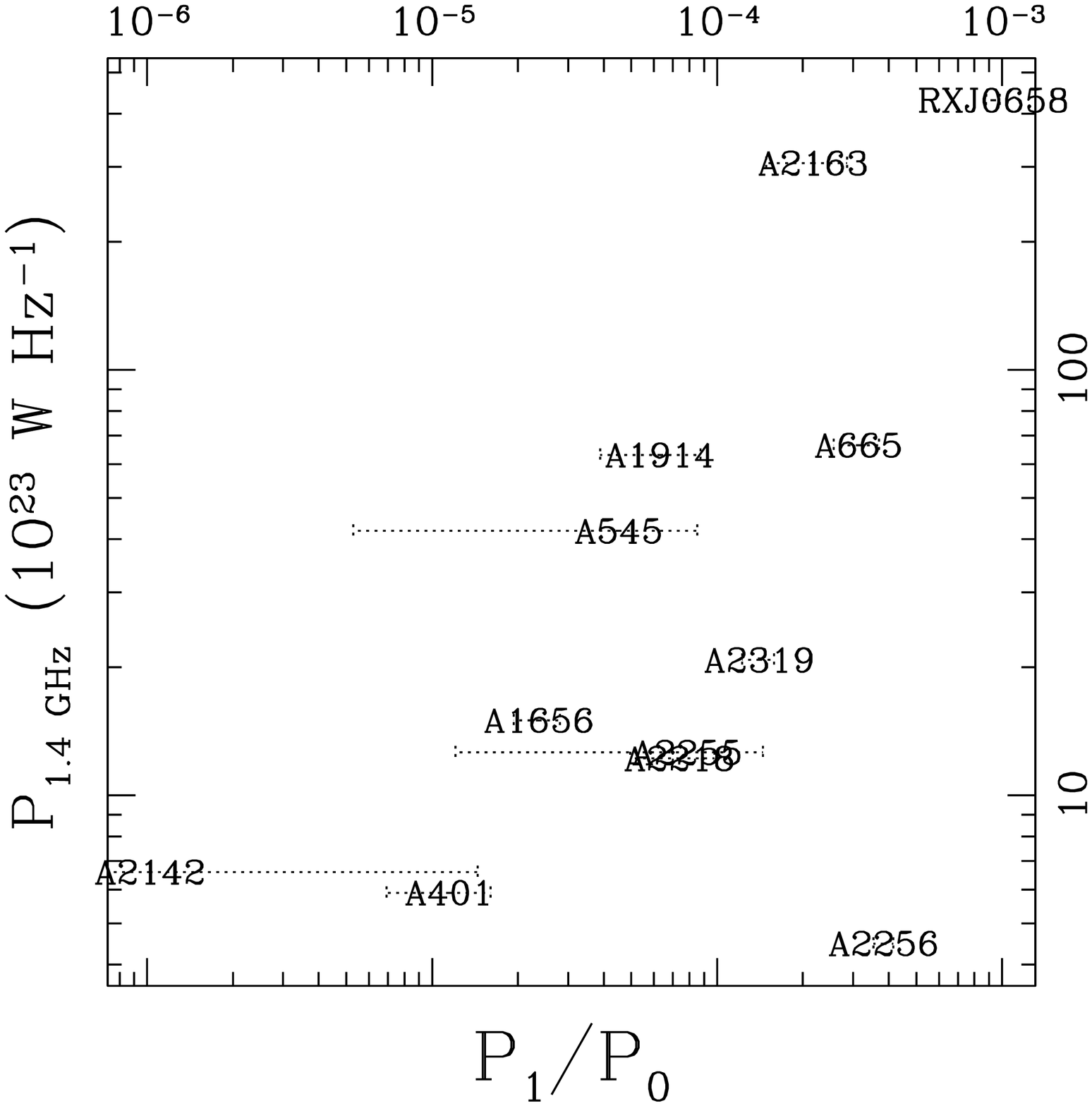,height=0.32\textheight}}}
\parbox{0.49\textwidth}{
\centerline{\psfig{figure=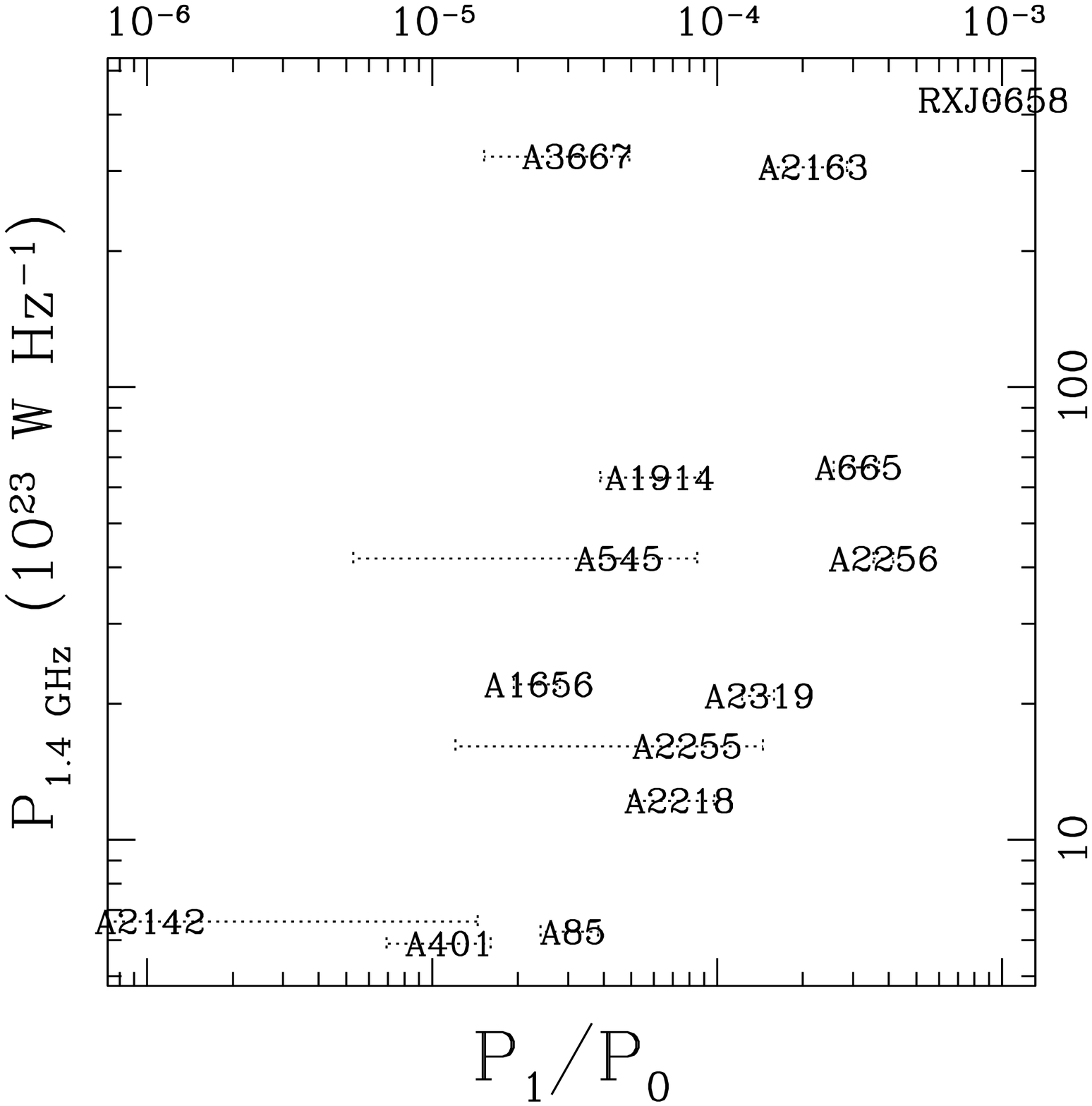,height=0.32\textheight}}}
\caption{\label{fig.radio} From Buote (2001): Radio power ($P_{1.4}$
-- 1.4 GHz rest frame) versus dipole power ratio ($P_1/P_0$) where
$P_{1.4}$ includes emission from (Left) only radio halos and (Right)
the total diffuse emission from halos and relic sources. The power
ratios are computed within a 0.5 Mpc aperture centered on the X-ray
emission peak with estimated $1\sigma$ errors shown. (Uncertainties on
$P_{1.4}$ are believed to be $\la 10\%$ and are not shown.) }
\end{figure}       

In Buote (2001) I used power ratios to provide the first quantitative
comparison of the dynamical states of clusters possessing radio halos.
A correlation between the 1.4 GHz power ($P_{1.4}$) of the radio halo
(or relic) and the magnitude of the dipole power ratio ($P_1/P_0$) was
discovered such that approximately $P_{1.4}\propto P_1/P_0$ (see
Figure \ref{fig.radio}). The $P_{1.4}-P_1/P_0$ correlation not only
confirms previous circumstantial evidence relating the presence of
radio halos to mergers but, more importantly, establishes for the
first time a quantitative relationship between the ``strength'' of
radio halos and relics ($P_{1.4}$) and the ``strength'' of mergers
($P_1/P_0$); i.e., the strongest radio halos appear only in those
clusters currently experiencing the largest departures from a
virialized state. Moreover, in the $P_{1.4}-P_1/P_0$ plane both radio
halos and relics may be described consistently which provides new
evidence that both halos and relics are formed via mergers.  The
$P_{1.4}-P_1/P_0$ correlation supports the idea that shocks in the
X-ray gas generated by mergers of subclusters accelerate (or
re-accelerate) the relativistic particles responsible for the radio
emission.

From additional consideration of a small number of highly disturbed
clusters without radio halos detected at 1.4 GHz, and recalling that
radio halos are more common in clusters with high X-ray luminosity
(Giovannini, Tordi, \& Feretti 1999), I argued that radio halos form
preferentially in massive ($L_{\rm x}\ga 0.5 \times 10^{45}$ erg
s$^{-1}$) clusters experiencing violent mergers ($P_1/P_0\ga 0.5
\times 10^{-4}$) that have seriously disrupted the cluster core. The
association of radio halos with massive, large-$P_1/P_0$,
core-disrupted clusters is able to account for both the vital role of
mergers in accelerating the relativistic particles responsible for the
radio emission as well as the rare occurrence of radio halos in
cluster samples.

On average $P_1/P_0$ is expected to increase with increasing redshift
owing to the higher incidence of merging (Buote 1998) which would lead
to a higher incidence of radio halos. However, on average cluster
masses are lower at earlier times implying a lower incidence of radio
halos. Each of these factors is dependent on the assumed cosmology,
and future theoretical work is therefore required to establish whether
the abundance of radio halos (1) increases or decreases with redshift,
and (2) provides an interesting test of cosmological models.

\section{Temperature Substructure}
\label{temp}

The morphologies of X-ray images of clusters suggest that clusters
span a wide range of dynamical states and merger configurations.
During such violent mergers the gas should be shock-heated at various
locations between an infalling subcluster and the center of the
primary cluster. In contrast to the azimuthally symmetric temperature
profile expected of a relaxed system, two-dimensional temperature
variations both represent a necessary confirmation of the merger
picture obtained from images and also provide a complementary view of
the cluster dynamical state and merger history.

\subsection{X-Ray Temperature Maps}

\begin{figure}[ht]
\vskip 0.75cm
\centerline{\psfig{figure=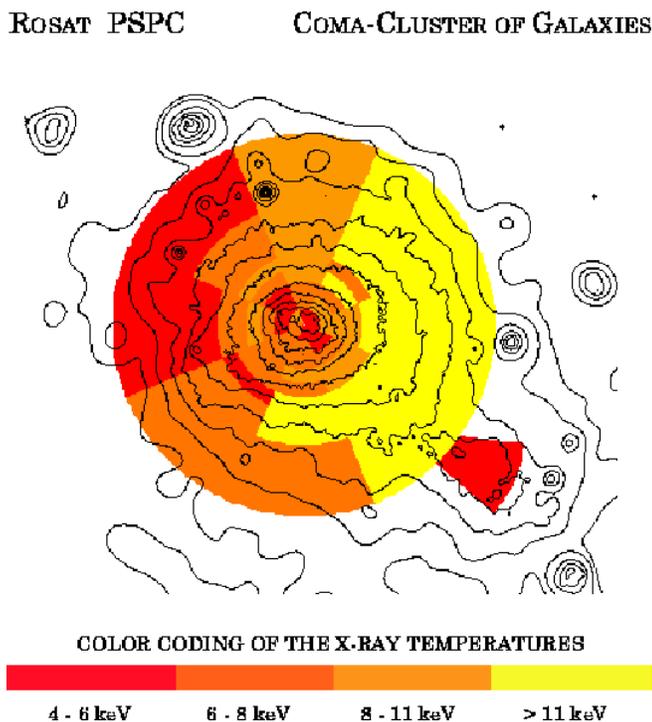,angle=0,height=0.49\textheight}}
\caption{\label{fig.rosat_tmap} {\sl ROSAT} temperature map of
the Coma cluster (Briel \& Henry 1997).}
\end{figure}

In the era before {\sl Chandra} and {\sl XMM} it was exceedingly
difficult to obtain accurate two-dimensional X-ray temperature maps of
clusters. The {\sl ROSAT} PSPC had sufficient spatial and spectral
resolution but its bandpass cut off sharply just beyond 2 keV. Since
massive clusters have temperatures above $\sim 5$ keV the temperatures
could not be constrained with any precision for all but a small number
of the brightest clusters. For these clusters the S/N was so high that
the data from the spectra below 2 keV managed to place interesting
constraints on the temperature.

For example, the {\sl ROSAT} temperature map of Coma (Briel \& Henry
1997) displayed in Figure \ref{fig.rosat_tmap} shows significant
temperature variations. The region of hotter gas in between the main
cluster and the NGC 4839 subcluster is consistent with shock heating
during the passage of the subcluster through the main cluster (e.g.,
Burns et al. 1994; Ishizaka \& Mineshige 1996). However, further
simulations are required to establish whether the subcluster is
currently falling in or has already passed through the main
cluster. As noted by Briel \& Henry (1997) if the subcluster already
passed though the main body then then it is unclear why the subcluster
still has retained its halo of hot gas.  Other {\sl ROSAT} temperature
maps of mergers display similar evidence for shock-heating (e.g.,
Briel \& Henry 1994; Henry \& Briel 1995, 1996; Ettori et al. 2000).

\begin{figure}[ht]
\parbox{0.49\textwidth}{
\centerline{\hskip -1cm \psfig{figure=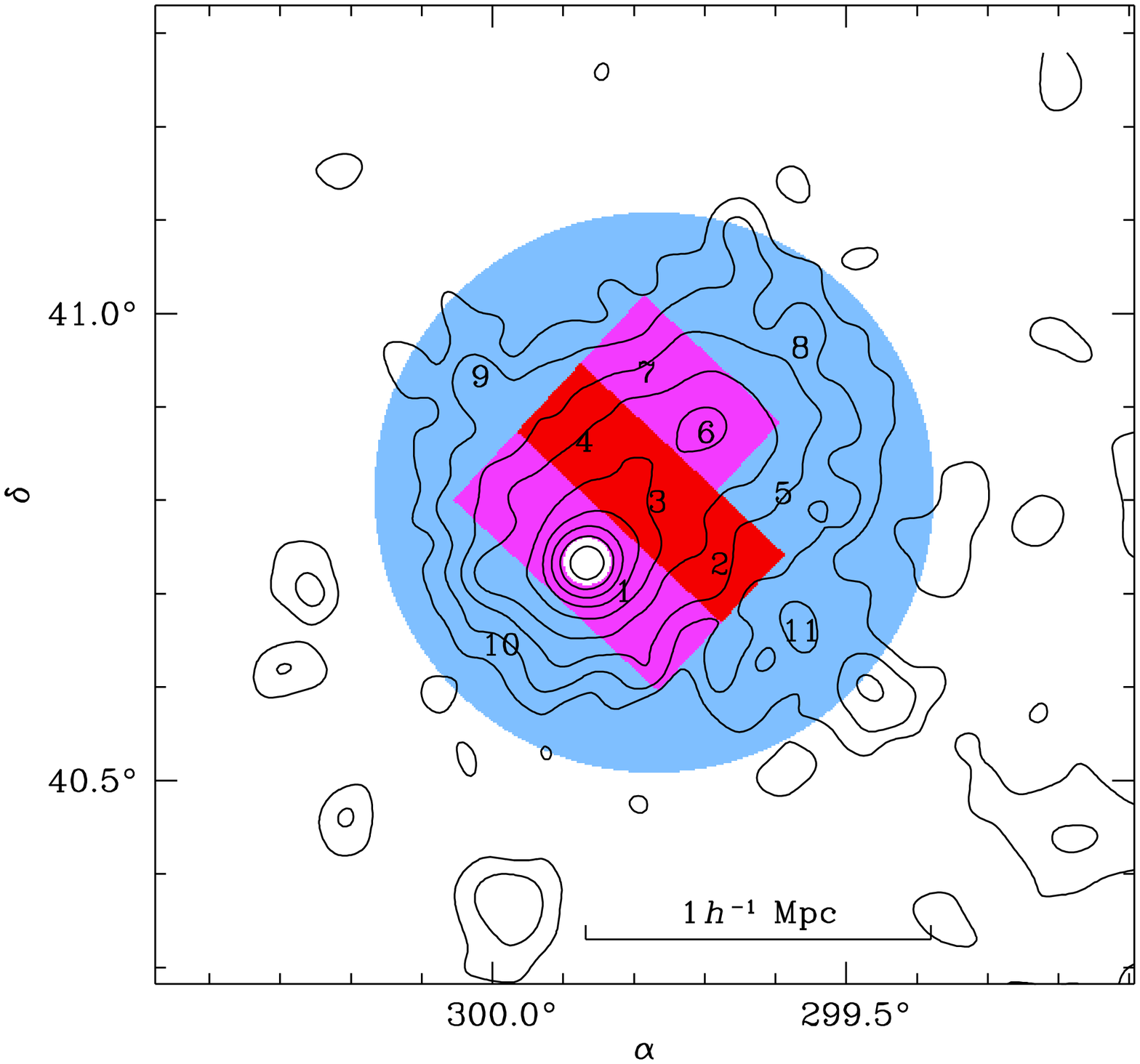,angle=0,height=0.35\textheight}}}
\parbox{0.49\textwidth}{
\centerline{\psfig{figure=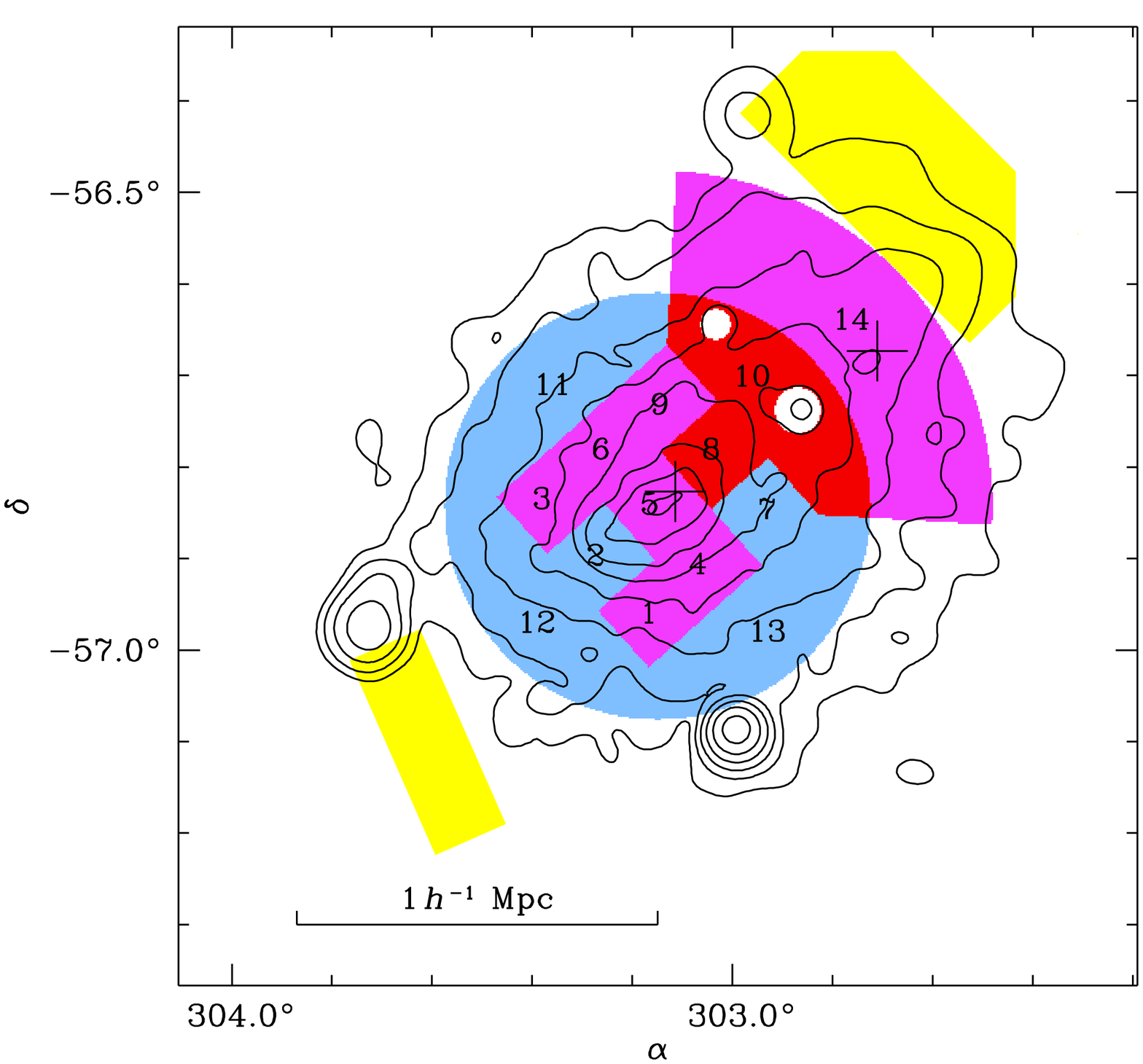,angle=0,height=0.35\textheight}}}
 
\vskip -0.25cm
 
\parbox{0.49\textwidth}{
\centerline{\hskip -0.25cm \psfig{figure=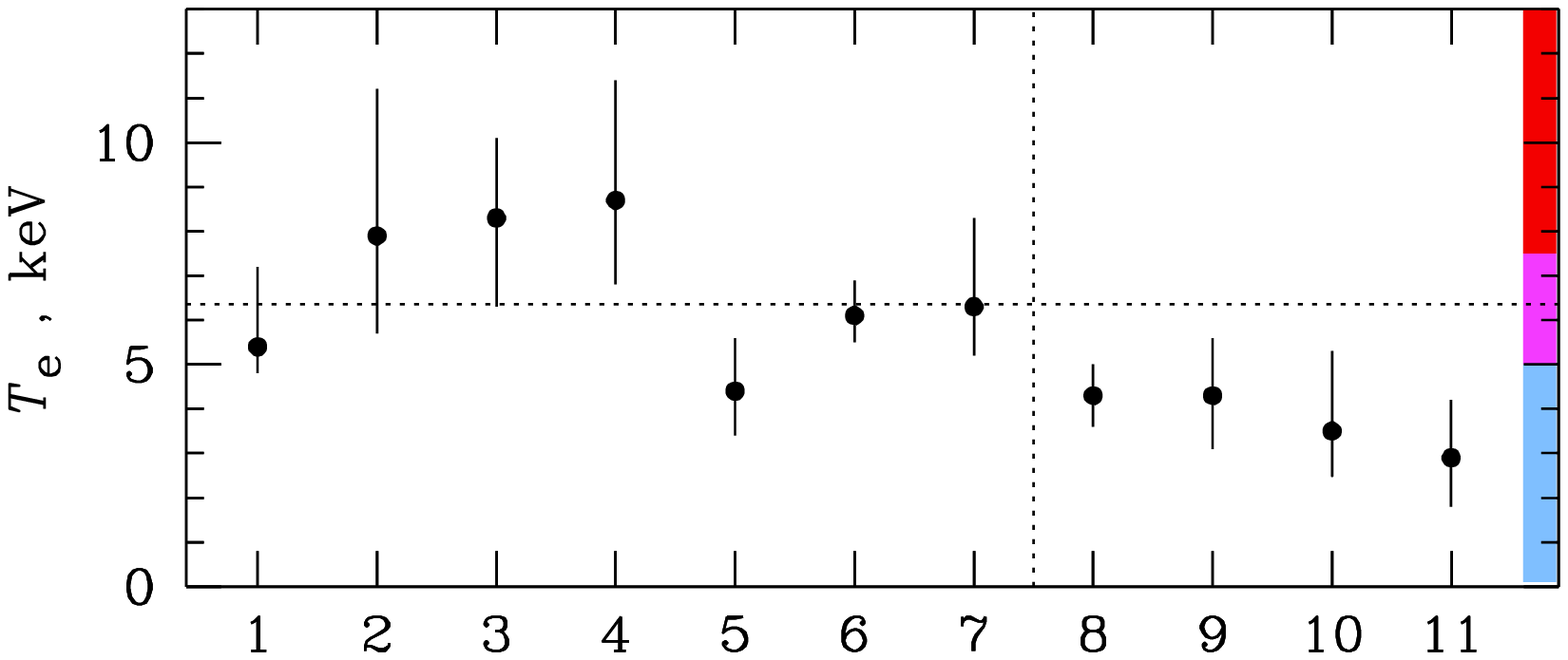,angle=0,height=0.16\textheight}}}
\parbox{0.49\textwidth}{
\centerline{\hskip 0.75cm \psfig{figure=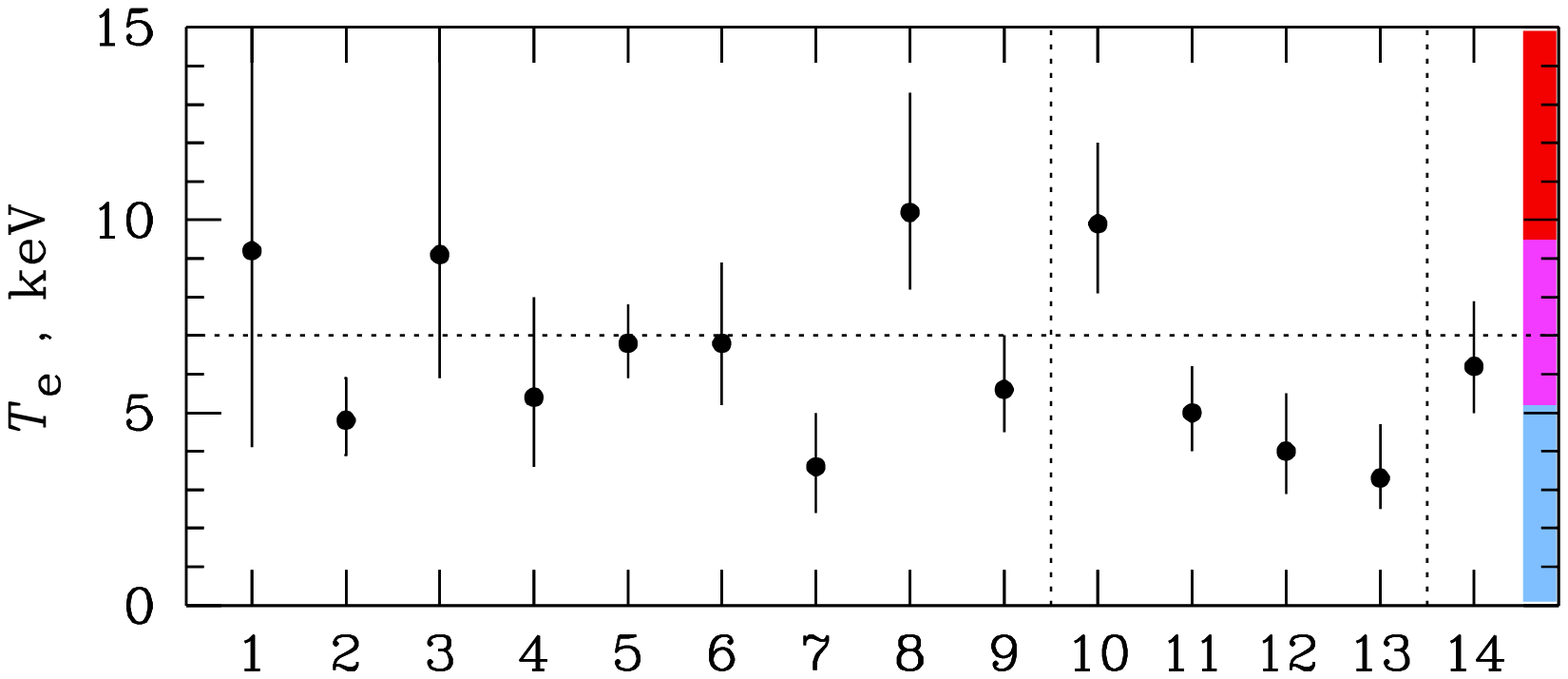,angle=0,height=0.16\textheight}}}
\caption{\label{fig.asca_tmap} {\sl ASCA} temperature maps of (Left)
Cygnus-A and (Right) A3667 from Markevitch et al. (1999).}
\end{figure}        

The qualitative features in the temperature maps derived from {\sl
ROSAT} were also found with data from the {\sl ASCA} satellite. The
higher energy resolution and larger bandpass (up to 10 keV) of {\sl
ASCA} provided a distinct advantage over {\sl ROSAT} studies, but the
poor spatial resolution $(\ga 1.5\arcmin$ FWHM) and highly energy
dependent point spread function (PSF) seriously hampered
two-dimensional spatial-spectral analysis. To obtain physical results
with {\sl ASCA} data the PSF needs to be incorporated into the
analysis.

When incorporating the PSF into modeling of the {\sl ASCA} data of
mergers two-dimensional temperature variations similar to those
obtained by {\sl ROSAT} are found. For example, in Figure
\ref{fig.asca_tmap} the results of the analysis of Cygnus-A and A3667
by Markevitch, Sarazin, \& Vikhlinin (1999) are shown. Although some of
the detailed results for a particular cluster differ between studies
using different deconvolution procedures, the basic idea that
non-azimuthal temperature variations exist in mergers seems to be
supported by most {\sl ASCA} and {\sl BeppoSAX} studies (e.g.,
Markevitch et al. 1998, 1999; Churazov et al. 1999; Donnelly et
al. 1999; Molendi et al. 1999; Shibata et al. 1999; de Grandi \&
Molendi 1999; Henriksen, Donnelly, \& Davis 2000; Iwasawa et
al. 2000).

Since there are some differences in the radial temperature profiles
obtained from {\sl ASCSA} data depending on the PSF deconvolution
procedure used (see White 1999 and Irwin \& Bregman 2000 and
references therein) the detailed temperature features obtained with
{\sl ASCA} do need to be confirmed with {\sl Chandra} and {\sl
XMM}. (As do those with {\sl BeppoSAX} because of its low spatial
resolution.)  Nevertheless, the overall trend of non-azimuthal
temperature structures and the shock-heating of the intra-cluster
medium are supported by the available {\sl ROSAT}, {\sl ASCA}, and
{\sl BeppoSAX} data.

\subsection{Quantitative Classification of Temperature Morphology}

\begin{figure}[ht]
\centerline{\psfig{figure=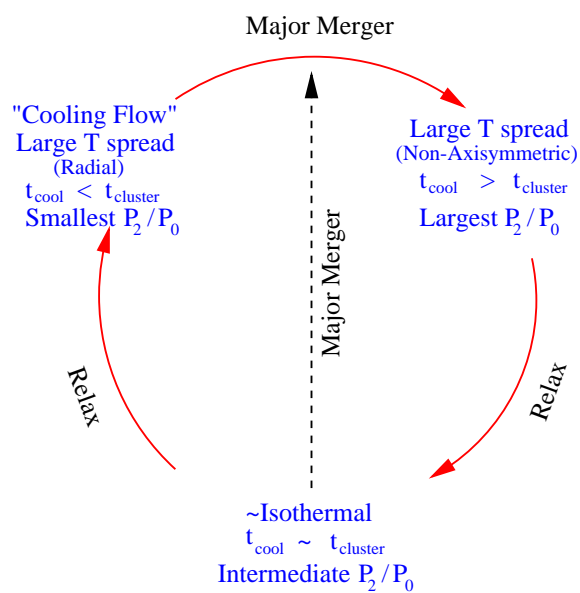,angle=0,height=0.4\textheight}}
\caption{\label{fig.temp} A possible description of the evolution of
the X-ray temperature structure and image morphology during the
formation and evolution of a cluster.}
\end{figure}

To obtain a more complete picture of the current dynamical states and
the merger histories of clusters the global morphological
classification of cluster images discussed in (\S \ref{global}) should
also incorporate the morphologies of X-ray temperature maps. In Figure
\ref{fig.temp} I show an idealized picture of how the temperature
morphology of a cluster might evolve during a merger. At early times
there is a large spread of temperatures distributed spatially in a
non-azimuthally symmetric fashion. At this time the cluster is far
from a virialized state. It possesses obvious substructure and a
disturbed spatial morphology quantified by, e.g., a large value for
the $P_2/P_0$ power ratio. The disturbed morphology implies there is
no cooling flow at this early time (see Figure \ref{fig.cf}), and the
cooling time ($t_{\rm cool}$) is longer than the cluster age ($t_{\rm
cluster}$) (e.g., Fabian 1994)

As the system relaxes image substructure and the spatial fluctuations
in the temperature are gradually erased until the system is
approximately isothermal and on the verge of establishing a cooling
flow. At this time $t_{\rm cool}\sim t_{\rm cluster}$ and there is
only a small amount of substructure (i.e., intermediate $P_2/P_0$
values -- see \S
\ref{global}).  If the cluster now experiences a major merger it will
begin again at the top of Figure \ref{fig.temp} with a lot of
temperature variations and image substructure. If instead the system
relaxes further without being disturbed then a cooling flow will
develop ($t_{\rm cool}< t_{\rm cluster}$) and the image substructure
should be mostly erased (smallest values of $P_2/P_0$). Although the
azimuthal temperature variations will also be erased, a radial
temperature gradient will be established where the temperature rises
from the center out to an approximately isothermal plateau.

Such radial temperature gradients are characteristic of cooling flows
(e.g., White 1999 and references therein). Whether the temperature
profile is caused by cooling gas or a two-phase medium (e.g., Ikebe et
al. 1997; Xu et al. 1998) is not important for the arguments
presented here. All that is required is that relaxed systems
(particularly those with cD galaxies) tend to have characteristic
radial temperature structure.

Therefore, for the merger scenario displayed by Figure \ref{fig.temp}
the amount of image substructure ($P_2/P_0$) falls continuously as the
cluster relaxes, but the overall spread in temperatures falls and then
rises again at late times. One possible way to quantify the
temperature morphology is with the multiphase strength (Buote,
Canizares, \& Fabian 1999) which essentially measures the width of the
differential emission measure, $\xi(T)$,
\begin{equation}
\sigma_\xi = { 1 \over 2 \langle T\rangle \xi_{\rm max} }
{\int_{T_{\rm min}}^{T_{\rm max}} \xi(T)\, dT},  \label{eqn.sigmaxi}       
\end{equation}
where $\xi_{\rm max}$ is the maximum value of $\xi(T)$ and $\langle
T\rangle$ is the emission-measure weighted value of $T$. This
statistic ignores the spatial information and is therefore intended as
a relatively crude measure of the temperature variations in a cluster
appropriate when the data do not allow precise temperature estimates
in small spatial regions. In such cases where the integrated cluster
spectrum is modeled with a simple cooling flow spectral model plus an
isothermal component then equation \ref{eqn.sigmaxi} is modified to
$f\sigma_{\xi}$, where $f$ is the relative fraction of the cooling
flow to the total emission measure (see section 5.2 of Buote et
al. 1999). A variation on this prescription using the breaks in
cooling flow mass deposition profiles has been used to determine the
``ages'' of some bright cooling flow clusters with {\sl ROSAT} (Allen
et al. 2000).

Joint consideration of $\sigma_{\xi}$ and $P_m/P_0$ should provide a
more precise indicator of the current cluster dynamical state and
merger configuration than $P_m/P_0$ alone. For high precision
temperature maps adding a first or second radial moment to equation
(\ref{eqn.sigmaxi}) may be sufficient to capture the spatial
dependences accounted for in the scenario represented in Figure
\ref{fig.temp}.

Finally, the scenario described by Figure \ref{fig.temp} will be
complicated if there are important dynamical contributions from
non-thermal processes such as AGN feedback (e.g., Owen et
al. 2001). Empirical studies of the spatial and spectral morphologies
of a large number of clusters using the improved instruments on {\sl
Chandra} and {\sl XMM} will help to elucidate the importance of these
and other process associated with cluster formation and evolution.

\section{Conclusions}

X-ray images of clusters obtained by {\sl Einstein} and {\sl ROSAT}
have established that substructure and merging are common in nearby
galaxy clusters. This evidence is reinforced by the X-ray temperature
maps of a smaller number of bright clusters analyzed by {\sl ROSAT},
{\sl ASCA}, and {\sl BeppoSAX}. The study of substructure and
morphology has evolved beyond detection and visual classification to
that of quantitative morphological statistics that probe the dynamical
states and and the power spectrum of density fluctuations.

Unfortunately, the present status of cosmological studies of cluster
morphologies is ambiguous. Although theoretical studies agree that
cluster morphologies are sensitive to the cosmology (particularly to
$\Omega_0$ and $P(k)$), the nature of the agreement and the
relationship to observations have been often in conflict. It is
difficult to interpret these disagreements because all of the N-body
simulations applied to this problem have been inadequate. Large
volume, high-resolution gas dynamical N-body simulations are required
to obtain definitive answers. A larger observational sample of cluster
morphologies with higher S/N data is also needed.

A quantitative connection between cluster mergers and the formation of
radio halos has now been established. The strength of a merger
indicated by the dipole power ratio ($P_1/P_0$) is approximately
proportional to the power of the radio halo. Radio halos form
preferentially in mergers of massive clusters with large values of
$P_1/P_0$ where the merger has proceeded fully into the core of the
cluster. Larger samples are needed to understand the relative
importance of the mass and $P_1/P_0$ on the strength of the radio halo
and to clarify the connection between the formation of radio halos and
relics.

\begin{acknowledgments}
I am grateful to the editors for the invitation to provide this
review. I thank J. Tsai and G. Xu for previous collaboration on
studies of cluster morphologies. I also thank the IAU, AAS, and a {\sl
Chandra} Fellowship for travel assistance to the IAU meeting in
Manchester in August, 2001 where I presented the material that formed
the basis for this review.
\end{acknowledgments}

\end{document}